\documentclass[12pt]{article}
\usepackage{graphicx,amsmath}
\usepackage{units}

\newlength{\dinwidth}
\newlength{\dinmargin}
\renewcommand{\vec}[1]{\boldsymbol{#1}}

\def\lapproxeq{\lower .7ex\hbox{$\;\stackrel{\textstyle                                                    
<}{\sim}\;$}}                                                    
\def\gapproxeq{\lower .7ex\hbox{$\;\stackrel{\textstyle                                                    
>}{\sim}\;$}}                                                    
\def\be{\begin{equation}}                                                    
\def\ee{\end{equation}}                                                    
\def\bea{\begin{eqnarray}}                                                    
\def\eea{\end{eqnarray}}
\def\b{\vec{b}}
\def\bb{\vec{b}'}

\def\k{\vec{k}_t} 
\def\kk{\vec{k}'_t} 
\def\GeV{\rm GeV}

\def\sh{\hat s}
\def\sh2{{\hat s}^2}
\begin{document}                                                    
\titlepage                                                    
\begin{flushright}                                                    
IPPP/11/10  \\
DCPT/11/20 \\                                                    
\today \\                                                    
\end{flushright} 
\vspace*{0.5cm}
\begin{center}                                                    
{\Large \bf High-energy strong interactions: from `hard' to `soft'}\\

\vspace*{1cm}
                                                   
M.G. Ryskin$^{a,b}$, A.D. Martin$^a$ and V.A. Khoze$^{a,b}$ \\                                                    
                                                   
\vspace*{0.5cm}                                                    
$^a$ Institute for Particle Physics Phenomenology, University of Durham, Durham, DH1 3LE \\                                                   
$^b$ Petersburg Nuclear Physics Institute, Gatchina, St.~Petersburg, 188300, Russia

\vspace*{1cm}                                                    
                                                    
\begin{abstract}                                                    
We discuss the qualitative features of the recent data on multiparticle production observed at the LHC. The tolerable agreement with Monte Carlos based on LO DGLAP evolution indicates that there is no qualitative difference between `hard' and `soft' interactions; and that a perturbative QCD approach may be extended into the soft domain. However, in order to describe the data, these Monte Carlos need an additional infrared cutoff $k_{\rm min}$ with a value $k_{\rm min} \sim 2-3$ GeV which is not small, and which increases with collider energy. Here we explain the physical origin of the large $k_{\rm min}$. Using an alternative model
 which matches the `soft' high-energy hadron interactions smoothly on to perturbative QCD at small $x$, 
we demonstrate that this effective cutoff $k_{\rm min}$ is actually due to the strong absorption of low $k_t$ partons. 
 The model embodies the main features of the BFKL approach, including the diffusion in transverse momenta, ln$k_t$, and an intercept consistent with resummed next-to-leading log corrections. Moreover, the model uses a two-channel eikonal framework, and includes the contributions from  the  multi-Pomeron exchange diagrams, both non-enhanced and enhanced. The values of a small number of physically-motivated parameters are chosen to reproduce the available total, elastic and proton dissociation cross section (pre-LHC) data. Predictions are made for the LHC, and the relevance to ultra-high-energy cosmic rays is briefly discussed. The low $x$ inclusive integrated gluon PDF, and the diffractive gluon PDF, are calculated in this framework, using the parameters which describe the high-energy $pp$ and $p\bar{p}$ `{\it soft}' data. Comparison with the PDFs obtained from the global parton analyses of deep inelastic and related {\it hard} scattering data, and from diffractive deep inelastic data looks encouraging.

\end{abstract}                                                        
\vspace*{0.5cm}                                                    
                                                    
\end{center}                                                    
                                                    
\section {Introduction}
 The general-purpose Monte Carlo generators \cite{gmc}, like PYTHIA or HERWIG, 
 describe the inclusive spectra observed in hadron collider experiments
assuming that there is some relatively hard parton-parton interaction in the central region, supplemented by the secondaries produced in  backward evolution from the hard matrix element to the incoming protons.
 It turns out that in order to reproduce the data, the infrared 
 cutoff, $k_{\rm min}$, in hard matrix element should increase with collider 
 energy reaching a value of about 3 GeV for $\sqrt s=7$ TeV 
(as compared to 2.15 GeV at the Tevatron energy)\footnote{The numbers correspond to PYTHIA 8.1 \cite{P81}.}. Such a large value of $k_{\rm min}$ should be explained within a perturbative QCD framework.

The most natural possibility is to say that in the low $x$ region,
relevant for high-energy collisions, the probability of rescattering,
and/or additional secondary interactions, becomes large on account of the high parton density. Then the corresponding absorptive corrections
(which are driven by a cross section $\sigma^{\rm abs}\propto 1/k^2_t$)
suppress the low $k_t$ contribution.

Here we study this possibility at a quantitative level. We consider a model based on the conventional perturbative QCD in the high $k_t$ domain, which includes the main features of 
the BFKL approach \cite{book} in a simplified form.
In this way we describe the structure of the `bare' Pomeron. Note 
 that this QCD Pomeron 
has an internal variable: the transverse momentum, $k_t$, of the intermediate partons.  In order to extend the description into the lower $k_t$  domain we include multi-Pomeron contributions
written in terms of Reggeon Field Theory (RFT) \cite{RFT}.  Recall that RFT offers the description of high-energy `soft' hadron-hadron interactions \cite{bkk}. Thus we achieve a smooth transition from the perturbative QCD to the `soft' domain. That is, we construct a model
which describes both pure `soft'
and `hard' phenomena within a common unified framework and with the same set of parameters. In other words, the model describes all facets of high-energy hadron interactions on the same footing.

\section{LHC data versus Monte Carlo predictions}
The inclusive single particle ($p_t,\eta$) distributions measured at the LHC  are in broad 
agreement with the Monte Carlo (MC) predictions, but there are significant discrepancies.  In the most popular MCs, 
like PYTHIA, even the soft interaction is simulated by a `hard' 
subprocess (say, $gg\to gg$,...) with 
\be
k_t~>~k_{\rm min}~=~k_0(s/s_0)^a,
\label{eq:1}
\ee 
accompanied by DGLAP evolution up to the scale, $k_t$, 
corresponding to the hard subprocess; plus hadronization at the final stage. The parametric form of $k_{\rm min}$ is tuned to the data. Leading order (LO) evolution and LO PDFs are 
used. However, since the model accounts for exact energy-momentum 
conservation, the results include some NLO effects.

\begin{figure} [t]
\begin{center}
\includegraphics[height=8cm]{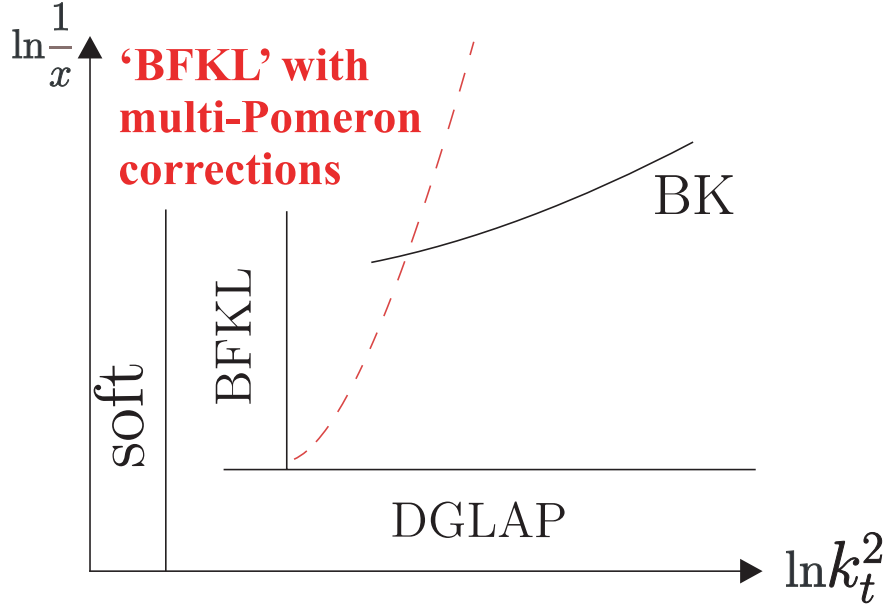}
\caption{\sf  The theoretical formalisms appropriate to the various domains are indicated. As long as $x$ is not too small, we have a well justified theory based on the perturbative
 QCD and DGLAP evolution.  For smaller $x$ we need to resum the 
 terms enhanced by the large value of $\ln 1/x$ inside  DGLAP 
 evolution, and to account for the 
 semi-enhanced absorptive corrections generated by the 
 Balitsky-Kovchegov (BK) equation \cite{BK}.  Above the `BK' curve more complicated multi-Pomeron diagrams enter and we cannot justify the result obtained by summation of `fan' diagrams only, that is, the results based on the BK-equation.
However, here we extend the partonic ladder structure of the Pomeron (generated by BFKL-like evolution in rapidity) to allow for a full set of multi-Pomeron exchange diagrams, and obtain a model which is applicable in the low $k_t$ region bounded by the dashed curve. Moreover, in principle, it is possible to use a more precise evolution kernel, which accounts for both BFKL and DGLAP logarithms, to cover the whole $k_t$ region in our approach. 
}
\label{fig:kin}
\end{center}
\end{figure}

We now turn to the discrepancies. In the low $p_t<0.5$ GeV region the high-energy data exceed the 
MC predictions by, for example, up to 40-50\% at $p_t\simeq 100$ MeV. To summarize, the data 
\cite{atlas,cms} for the
$q_t$-integrated particle density $dN/d\eta$ turn out to be larger, 
while the mean $p_t$ ($\langle p_t \rangle$) is smaller, than the predictions of the MCs. These observations indicate the presence 
 of the configurations which 
violate the strong $k_t$ ordering, which is characteristic of LO DGLAP. Rather, at high energies, where we enter the low $x$ domain, 
BFKL-like contributions are important, see Fig.\ref{fig:kin}. That is, contributions which are ordered in $x$ but not in $k_t$.

\subsection{The `old' Regge description of `soft' data}
Before we introduce the BFKL amplitude, which sums up diagrams with so-called multi-Regge kinematics, we first recall the main features of the `old' Regge description of soft high-energy hadron interactions,
see, for example, \cite{bkk,regge}.
In this approach, the high-energy interaction is described by Pomeron exchange,
By `Pomeron' we mean the rightmost singularity in the complex angular momentum or $j$ plane. In the simplest case, the Pomeron is a 
pole in $j$-plane, which originates from the sum of ladder-type diagrams in which the transverse momenta of the particles are assumed to be limited, and not to grow with energy.

Indeed,  up to Tevatron energies the {\it elastic} amplitudes, are well described with a Pomeron which is a single pole at \cite{DL} 
\be
\alpha_P^{\rm eff}(t) \equiv  j=1+\epsilon+\alpha't, ~~~~~~~ {\rm with} ~~\epsilon=0.08~~ {\rm and} ~~\alpha'=0.25~ \GeV^{-2}.
\label{eq:DL}
\ee
The reason for the superscript `effective' will become clear in a moment.
On the other hand, in this one-pole approximation the {\it single-particle inclusive} cross section grows in the same way as the inelastic cross section, $d\sigma/dy d^2p_t=V_{\rm in}(p_t^2)\sigma_{\rm inel}$, see
\cite{M-K}.
 In this case the particle density,
\be
 \frac{dN}{dy dp^2_t}~ =~\frac{1}{\sigma_{\rm inel}} \frac{d\sigma}{dy dp^2_t},
\ee
 does not depend on energy, contrary to the data.

The puzzle is solved by including the contribution of the multi-Pomeron exchange diagrams.
Due to the cancellation\footnote{This cancellation can be easily seen by applying the AGK cutting rules \cite{AGK} to the (eikonal) multi-Pomeron diagrams. 
{\it Eikonal} diagrams correspond to Pomerons exchanged
 between the colliding protons. Later we will discuss so-called {\it enhanced} multi-Pomeron diagrams.} it turns out that the single-particle 
 inclusive cross section is still described by one Pomeron exchange,
\be
d\sigma/dy dp^2_t\propto s^\Delta
\ee
whereas the growth of inelastic cross section is reduced by the contribution of the multi-Pomeron diagrams. In other words, $\epsilon$ in (\ref{eq:DL}), is an effective power\footnote{The `effective' pole form, (\ref{eq:DL}), only provides a reasonable description up to Tevatron energies. The elastic amplitude, $T_{\rm el}(s,b)$, already exceeds the black disc limit at small impact parameters for $\sqrt{s} \gapproxeq 5$ TeV,
see, for example, \cite{cer}.}, which is smaller than $\Delta$ of the original `bare' Pomeron pole: 
\be
\alpha_P^{\rm bare}(t)=1+\Delta+\alpha_P^{\prime~\rm bare}(t).
\label{eq:Delta}
\ee

In the present Monte Carlos these phenomena are included as the possibility of Multiple Interactions (MI) when a few pairs of incoming partons interact simultaneously, producing  independent\footnote{Modulo the constraint of the conservation of the overall energy and momentum.} chains of parton showers. For each chain the evolution is convoluted with the corresponding `hard'
subprocess.

\subsection{Implications of the LHC data}

In the energy interval covered by the current LHC data, $0.9<\sqrt{s}< 7$ TeV, the particle density in the central region, $\eta=0$, grows as \cite{atlas,cms}
\be
dN/d\eta\sim s^{0.12}.
\ee
  Taken together with the behaviour of the inelastic cross 
section $\sigma_{\rm inel}\sim s^{0.08}$, it means that the two-particle irreducible amplitude described by the `bare' Pomeron pole, which according to AGK rules also specifies the inclusive cross section, leads to the result
\be 
\frac{d\sigma}{dy}=\sigma_{\rm inel}\frac{ dN}{dy} \sim s^\Delta~~~~~~{\rm with}~~\Delta\simeq 0.2.
\ee

Simultaneously, the mean transverse momentum of secondaries observed at the LHC \cite{cms} is found to have the behaviour
\be
\langle p_t \rangle \propto s^{0.05}.
\ee
In other words, the particle multiplicity, $dN/dy$, in the central region grows mainly due to the population of a larger phase space ($\propto \langle p_t \rangle^2$) in the transverse-momentum distribution, while at very low $p_t$ the particle density $dN/dy dp^2_t$ is close to saturation. It should be emphasized that the energy dependence of mean $\langle p_t \rangle$ is beyond description by `old' Regge theory.


\subsection{Diffusion in $\ln k_t$}
In Monte Carlos, based on LO DGLAP evolution, the transverse momenta of intermediate partons monotonically increase starting from the proton up to the matrix element of `hard' subprocess taking place somewhere in the central region. For the events with a really hard subprocess (characterised by a large scale) such configurations, ordered in $k_t$, indeed give the dominant contribution. However, in a normal inelastic event (with no large scale) there is no reason for ordering in $k_t$.

In particular, within BFKL evolution\footnote{The evolution is started from some relatively large scale in order to justify pQCD and to neglect confinement. An example is an onium-onium interaction
\cite{BL}.} the transverse momentum can increase or decrease with equal probability, leading to  {\it diffusion}
in $\ln k_t$ \cite{BFKL86}. Thus, for the normal inelastic event, it would be natural to expect a large contribution from diagrams corresponding to `partonic chains' not ordered in $p_t$.

It is {\it surprising} that the recent high-energy LHC data on inclusive single-particle distributions are reasonably well described by the DGLAP-based Monte Carlos\footnote{Another surprise is that the reasonable description was obtained using the LO gluon PDF which grows with $1/x$ much steeper than that in NLO case. Recall that experimentally the low $x$ behaviour is fixed mainly by deep inelastic data where the heavy photon interacts with a quark. The absence of the $1/z$ singularity in the LO quark-quark splitting function, $P_{qq}(z)$, is compensated by a steeper $1/x$ behaviour of the LO gluon distribution.
At first sight, the NLO PDF should be more relevant for the inclusive $pp\to X$ cross sections measured in the central region where the gluon-gluon interactions dominate.}. Partly, this may be explained by the number of
parameters (which are not fixed by theory)  used to tune the Monte Carlos.

On the other hand, the fact that a rather large and increasing with energy  infrared cutoff, (\ref{eq:1}), is required to describe the data,
may indicate the suppression of the low $k_t$ contributions which 
 enforces the increase of transverse momenta along the parton chain in going from the protons to the central plateau.

\subsection{Low $p_t$ suppression and enhanced multi-Pomeron diagrams \label{sec:hyp}}
As mentioned in the Introduction, the suppression of low $k_t$ partons was actually expected. The absorptive cross section has the form $\sigma^{\rm abs}\propto 1/k^2_t$. Therefore
the absorptive effects stem the growth of `wee' parton densities at low $k_t$, while at larger $k_t$ the growth with $1/x$ is continued. In terms of Regge Field Theory (RFT) the absorption of intermediate partons is described by so-called `enhanced' multi-Pomeron diagrams.

\begin{figure} 
\begin{center}
\includegraphics[height=4cm]{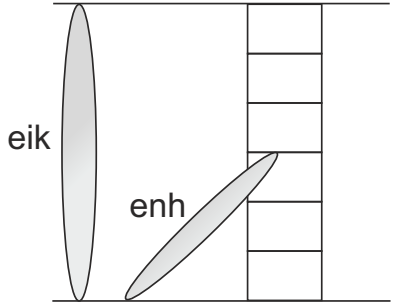}
\caption{\sf Eikonal and enhanced multi-Pomeron contributions}
\label{fig:ee}
\end{center}
\end{figure}

\begin{figure}
\begin{center}
\includegraphics[height=3cm]{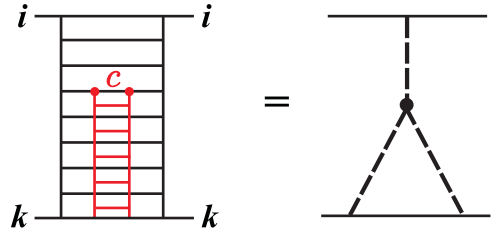}
\caption{\sf The triple-Pomeron diagram, together with its ladder structure}
\label{fig:2lad}
\end{center}
\end{figure}

So far, we have discussed only {\it eikonal} diagrams, with multi-Pomerons exchanged between the colliding protons. However, there is also the possibility that one or more Pomerons may couple to an intermediate parton. We then have a so-called {\it (semi)enhanced} multi-Pomeron diagram. The eikonal and enhanced diagrams are sketched in Fig.~\ref{fig:ee}. The simplest enhanced diagram corresponds to the triple-Pomeron contribution shown in Fig.~\ref{fig:2lad}.  Recall, that this diagram gives the leading contribution to the cross section for a proton dissociating into a high-mass system. The triple-Pomeron coupling\footnote{$g_{3P}\equiv g^2_1=g^1_2$ in the notation introduced below.}, $g_{3P}$, is found \cite{LKMR,deOliv} to be rather small, $g_{3P}/g_N \simeq 0.2$, where $g_N$ is the Pomeron-proton coupling. Nevertheless, the final effect of the triple-Pomeron diagrams is rather strong at high energies, since the contribution of each diagram is {\it enhanced} by the large rapidity interval that is available for the position of the vertex.
Indeed the enhanced diagrams lead to increasing amounts of high-mass dissociation with increasing collider energy.

The simplest models contain only multi-Pomeron enhanced diagrams generated by 
triple-Pomeron vertices. On the other hand, such models lead to cross sections which asymptotically decrease with energy \cite{bor}. We note that there is no reason to assume that all the more complicated multi-Pomeron vertices, $g^n_m$ with $n+m>3$, are exactly zero. In fact it is more reasonable to include these contributions. However, the values of the vertices, $g^n_m$, which couple $n$ to $m$ Pomerons, are not known for $n+m>3$. In this paper we study the two most physically reasonable hypotheses.
We formulate the hypotheses about the $g^n_m$ behaviour in terms of conventional soft RFT in which the transverse momenta of partons are limited and where, for the moment, we omit the variable $k_t$. 
 
The first hypothesis \cite{KMRns} is that
\be
 {\rm (a)}~~~ g^n_m=\Delta (n m/2) (g_N\lambda)^{m+n-2}.
\label{eq:(a)}
\ee
This corresponds to a suppression 
\be
\exp(-\lambda\Omega)
\label{sap-a}
\ee
of each intermediate parton; like the absorption of the incoming beam particle in a nuclear target with the opacity $\Omega$.
The second hypothesis is that
\be
{\rm (b)} ~~~g^n_m=G_{3P}(g_N\lambda)^{m+n-2} ~~~{\rm with}~~ G_{3P}=\Delta,
\label{eq:(b)}
\ee
where the last condition ensures that we are in the `critical' Pomeron regime\footnote{More details are given in Section \ref{sec:ost}.}.
After the enhanced screening the `renormalized' intercept (i.e. the position of the singularity in the $j$-plane) of the Pomeron is $\alpha^{\rm ren}(0)=1$. 
In this scenario the suppression factor at each intermediate parton vertex takes the form
\be
\frac{1-\exp(-\lambda\Omega)}{\lambda\Omega},
\label{sap-b}
\ee
instead of the (\ref{sap-a}).

For a smaller value of $G_{3P}<\Delta$ the screened intercept $\alpha^{\rm ren}(0)>1$. That is enhanced absorption is not sufficient to stop the power growth of the cross section. Only the eikonal diagrams restore unitarity. Actually this, `supercritical' regime is like an eikonal model with the internal structure of the Pomeron slightly modified.
On the other hand, a larger value of $G_{3P}$ leads to a decreasing cross section which dies out with energy.
We regard these last two regimes as unrealistic, and do not consider them further. That is we focus on the
`critical' regime which is provided automatically in case (a), or with a specific value of $G_{3P}$ in case (b).
Further discussion is given in Section \ref{sec:ost}.

Next, we would like to trace the smooth transition from the `soft' low $p_t$ domain to the `semi-hard' region of relatively large $p_t$.

\section{Matching the `soft' and `hard' contributions \label{sec:A3}}

There is evidence of a smooth transition from the `soft' Pomeron which describes `soft' data to the perturbative QCD Pomeron describing the semi-hard region of relatively large $p_t$, see subsection \ref{sec:evid}. This opens the way to link the description of high-energy soft interactions to the perturbative very low $x$, ~$p_t\sim$ few GeV domain, a region heavily populated by LHC data, see Fig.~\ref{fig:kin}. We sketch how this is done in subsection \ref{sec:proc} and give more details in the Appendix.

\subsection{Smooth transition between the hard and soft regimes  \label{sec:evid}}

There are phenomenological hints that at large distances the ``soft'' Pomeron should have qualitatively similar structure as the ``hard'' (QCD) Pomeron,
see \cite{shuv}. Indeed,
 first, no irregularity is observed in the HERA data in the transition region, $Q^2\sim 0.3 - 2 ~{\rm GeV}^2$, between the `soft' and `hard' interaction domains; the data are smooth throughout this region. In particular, the observed cross sections for vector meson production, $\gamma p \to Vp$, are consistent with a `soft' Pomeron of intercept $\alpha_P(0)\sim 1.1$ at low ($Q^2+M_V^2$), rising smoothly to $\alpha_P(0)\sim 1.3$ at large ($Q^2+M_V^2$).
Next, a small slope $\alpha'_P \lapproxeq 0.05 ~{\rm GeV}^{-2}$ 
 of the bare Pomeron trajectory, is obtained in global analyses of all available soft high-energy data, after
  accounting for absorptive corrections and secondary Reggeon contributions \cite{KMRnns1,GLMM}.  This indicates that 
the typical values of $k_t$ inside the Pomeron amplitude are relatively large ($\alpha'\propto 1/k^2_t$).
 Finally, recent `soft' model data analyses \cite{KMRnns1,GLMM} which account for the enhanced absorptive effects find an intercept of the initial, bare Pomeron $\Delta=\alpha_P(0)-1\simeq 0.3$ close
  to the intercept of the BFKL Pomeron after the NLL corrections are resummed \cite{bfklresum,kmrsre}.
Thus it looks reasonable to assume that in the soft domain we deal
 with the same perturbative QCD Pomeron;  at least, there is a smooth transition from the soft to the hard Pomeron.

The net effect is that the {\it bare} hard Pomeron, with a trajectory with intercept $\Delta \equiv \alpha_P(0)-1 \simeq 0.3$ and small slope $\alpha '$, is subject to increasing absorptive effects as we go to smaller $k_t$ which allow it to smoothly match on to the attributes of the {\it soft} Pomeron. In the {\it limited} energy interval up to the Tevatron energy, some of these attributes (specifically those related to the elastic amplitude) can be mimicked or approximated by an {\it effective} Pomeron pole with trajectory intercept $\Delta_{\rm eff} \equiv \alpha^{\rm eff}_P(0)-1 \simeq 0.08$ and slope $\alpha'_{\rm eff}=0.25~{\rm GeV}^{-2}$ \cite{DL}.

\subsection{Outline of the procedure to link the hard and soft regimes  \label{sec:proc}}

\begin{figure} 
\begin{center}
\includegraphics[height=5cm]{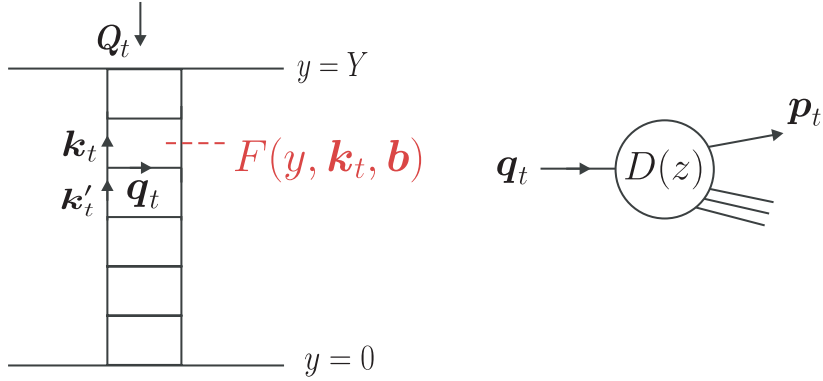}
\caption{\sf The ladder structure of the {\it bare} Pomeron, $F(y,\k,\b)$, together with a symbolic sketch of the hadronisation process, cf (\ref{eq:D}). This figure shows the symbols for the various transverse momenta used in the text.}
\label{fig:Pladder}
\end{center}
\end{figure}

We start with the partonic ladder structure of the Pomeron, $F(y,\k,\b)$, generated by BFKL-like evolution in rapidity, see Fig.~\ref{fig:Pladder}
\be
\frac{\partial F(y,\k,\b)}{\partial y}~=~\int d^2\bb \int \frac{d^2\kk}{\pi k^{'2}_t}~K(\k,\kk)~F(y,\kk,\bb)\delta(\b-\bb).
\label{eq:ev}
\ee
At each step of the evolution $\k$ and the impact parameter, $\b$, can be changed.   It is important to note that, in comparison with RFT, we now have an extra variable, $\k$, that is the transverse momentum of the intermediate parton, in addition to the usual two `soft' Regge variables: $y={\rm ln}(1/x)$ and the impact parameter, $\b$, which is conjugated to the transverse momentum, $Q_t$, transferred through the entire ladder.

Usually the evolution equation in perturbative QCD is written in terms of the amplitude integrated over the impact parameter $\b$. However here we have included the explicit $\b$ dependence, since then it is easier to calculate the effect of the multi-Pomeron contributions. In general the kernel $K$ depends on the difference $\b-\bb$. However, the $\b$ dependence is proportional to the slope $\alpha'$ of the bare Pomeron trajectory. Asymptotically the BFKL approach predicts $\alpha' \to 0$. Indeed, analyses of soft data find it to be very small, $\alpha'\lapproxeq 0.05 ~\GeV^2$, see \cite{KMRnns1,GLMM}.

. For simplicity, we neglect this variation. Then the only $\b$ dependence of $F$ comes from the starting distribution of the evolution, and not from the $\b$ dependence of $K$.

The sum of the non-enhanced eikonal
diagrams may be written in terms of the opacity $\Omega$. At each value of the impact parameter the result for the amplitude is
\be
T(\b)=1-\exp(-\Omega(\b)/2).
\label{eq:Tb}
\ee
Since at high energies the proton opacity $\Omega$ is proportional to the gluon density,
$F(y,\k,\b)$, unintegrated over the impact parameter  $\b$
it is convenient to write the linear evolution equation, (\ref{eq:ev}), in terms of $F$.  Note that $F$ depends not only on $\b$ and $y$ (as in soft RFT), or only on $y$ and $\k$ (as in pQCD), but on all three variables.

\begin{figure} 
\begin{center}
\includegraphics[height=3cm]{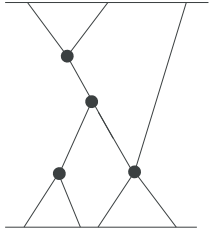}
\caption{\sf A typical enhanced multi-Pomeron exchange diagram}
\label{fig:coup}
\end{center}
\end{figure}

To include rescattering of the intermediate partons with the beam $i$ and target $k$ diffractive eigenstates, as in Fig~\ref{fig:coup}, we have to include in (\ref{eq:ev}) the absorptive factors (\ref{sap-a}) or (\ref{sap-b}), depending on one or another version of the model and to solve
two coupled evolution equations.
 One evolving up from the target $k$ at $y=0$, and one evolving down\footnote{The value of $Y_k$ accounts for the fact that at larger $k'_t$ a smaller rapidity interval is available for the evolution.} from the beam $i$ at $y'=Y_k-y=0$ with $Y_k={\rm ln}(s/k^{'2}_t)$,
\be
\frac{\partial F_k(y)}{\partial y}~=~\int \frac{d^2\kk}{\pi k^{'2}_t}~{\rm exp}(-\lambda[\Omega_k(y)+\Omega_i(y')]/2)~K(\k,\kk)~F_k(y).
\label{e15}
\ee
\be
\frac{\partial F_i(y')}{\partial y'}~=~\int \frac{d^2\kk}{\pi k^{'2}_t}~{\rm exp}(-\lambda[\Omega_i(y')+\Omega_k(y)]/2)~K(\k,\kk)~F_i(y'),
\label{e16}
\ee
where, for clarity, we have suppressed the $\k$ labels of the $F$'s and $\Omega$'s. Here we show the more familiar absorptive factors, exp$(-\lambda\Omega /2)$, of the amplitude at the intermediate parton vertices\footnote{Since we are dealing with the {\it amplitude} $F$, and not with the cross section, we use here $\exp(-\lambda\Omega/2)$ and not $\exp(-\lambda\Omega)$. } corresponding to hypothesis (a) of (\ref{eq:(a)}):
 in the case (b) it should be replaced by the form (\ref{sap-b}). The coupled evolution equations may be solved iteratively to give $F(y,\k,\b)$, for a whole range of fixed values of $\b$.

 The opacities which enter (\ref{e15},\ref{e16}) are given by the LO expression\footnote{The normalisation of this form has been adjusted such that $F(y,k_t,\b)$ is the doubly-unintegrated gluon distribution, as given in (\ref{eq:glut}) below.  We use the LO pQCD couplings with one-loop 
$\alpha_s$ ($n_f=3$, $\Lambda=150$ MeV) and allow for some renormalization due to NLO effects.} 
 \be
 \lambda\Omega_k(y,\k,\b)=\int_{k^2_t}N_c\pi^2\frac{dk^{'2}_t}{k^{'2}_t}\alpha_s(k^{'2}_t)F_k(y,k'_t,\b),
 \label{omeg}
 \ee
 where the integral over $k'_t$ accounts for the possibility of screening an 
 intermediate parton $c$ with transverse momentum $k_t$ by any additional ladder of smaller
 size\footnote{At a small $k'_t<k_t$ the gluon with a large Compton wave interact coherently with the colour charge of the whole (colour neutral) ladder. This contribution is negligible.}, that is with $k'_t>k_t$,
see \cite{theta}.
 
 To allow for low-mass dissociation, $p \to N^*,...$, we follow 
 Good-Walker \cite{GW} and introduce diffractive eigenstates, 
 $i,k$, which are those linear combinations of $p,N^*,...$ which 
 diagonalise the $T$ matrix and only undergo elastic-like 
 scattering. Thus we have to consider the interaction of the beam state $i$ with the target state $k$; correspondingly the opacities $\Omega$ in (\ref{e15},\ref{e16}) are marked by the indices $i$ and $k$. Two channels suffice \cite{KMRns}. 
 
 Recall that the kernel $K$ provides the possibility of evolution in both directions - the momentum $|\k|$ may be  larger or lower than $|\kk|$ with equal probability. On the other hand the absorptive factor $\exp(-\lambda\Omega(y,k_t))$ suppresses the production of a low $k_t$ partons. This acts like an infrared cutoff, $k_{\rm min}$, and makes the evolution asymmetric, leading to a larger probability of evolution with $k_t$ increasing; that is, in the usual DGLAP direction.
 
Besides this, we need to introduce a boundary condition at low $k_t$, say $k_t=q_0$, close to the confinement region. We assume that the parton density is zero for $k_t<q_0$. Moreover, to avoid possible double counting, and to fix the boundary
between the low- and high-mass dissociation, we introduce a threshold
$\Delta y=1.5$ in rapidity at which we start the evolution of
(\ref{e15},\ref{e16}). That is, we start the upwards evolution at $y=y_0=\Delta y=1.5$, and not at $y=0$, 
and the downwards evolution from  $y'=y_0=\Delta y=1.5$, and not from $y'=0$.   The relation between $y$ and $y'$ is 
\be
y'=Y_k-y   ~~~~~{\rm where}~~~~Y_k = \ln(s/k_t^2),
\label{eq:Yk}
\ee
recall $k_t>q_0$.  $y_0$ specifies the low- and high-mass dissociation regions, and high-mass dissociation takes place in a rapidity interval $\ln(s/k_t^2)-2\Delta y$. Low-mass dissociation is allowed for via the diffractive eigenstates, and high-mass dissociation is included via the enhanced multi-Pomeron exchange diagrams.

\subsection{The parameters of the model \label{sec:para}}
The idea is to study, in a semi-quantitative way,  the main features  of soft and semi-hard data in terms of a realistic model with just a {\it few} physically-motivated parameters, and {\it not} to perform a multi-parameter $\chi^2$-analysis of the data. In this way we hope that we can provide a better understanding of the physics which underlies the description of the data. Using the procedure outlined above, we start from a formulation in the larger $k_t$ (perturbative QCD) domain, and then extend the description into the lower $k_t$ region in order to reproduce the `soft' data.                                                                           

It is informative to gather together the small number of parameters, that we have introduced above (or have implied), that are to be tuned to describe all the gross features of the available `soft' high-energy data. There is basically one parameter (or sometimes two)  that is mainly responsible for each phenomena:
\begin{itemize}
\item $\Delta=\alpha_P(0)-1$ giving the intercept of the {\it bare} Pomeron trajectory, which controls the energy dependence.
In other words, the kernel $K$ is normalized to give a final intercept $\Delta$ for the bare Pomeron before absorptive effects are included.  The slope $\alpha'_P$ of the trajectory is set to zero, since it is found to be very small.
\item $d$ which provides a smooth transition between the logarithmic contributions with $k_t \ll k'_t$ and $k_t \gg k'_t$. It thus controls the BFKL-like diffusion in ln$k_t$.
\item  $~ N'$ which specifies the initial gluon density, and in this way fixes the Pomeron-proton coupling.
\item the parameters $c_1$ and $c_2$ which determine the proton radius and the shape of the corresponding form factor.
\item $\lambda$ ( or rather $f_{3p}$ of (\ref{eq:g3-bk}) below) which determines the strength of the triple- (and multi-) Pomeron couplings, which are constrained by data on high-mass diffractive dissociation.
\item $\gamma_i$ which specify the Good-Walker diffractive eigenstates \cite{GW}, and which are determined by low-mass diffractive
dissociation (the coupling of state $i$ to the Pomeron is proportional to $\gamma_i$).
\item $y_0$ which, in order to avoid double counting, separates low- and high-mass diffraction (the low mass dissociation is written in terms of the Good-Walker eigenstates while the high mass dissociaton is described in terms of multi-Pomeron (RFT) contributions).
\item $q_0$, the infrared cutoff, which together with $N'$, controls the absolute value of the bare one-Pomeron exchange cross section.
\end{itemize}

Except for the important introduction of an explicit treatment of the $\k$ dependence, all other features of the model are practically the same as described in our previous publications \cite{KMRns,KMRnns1}.

\subsection{Simultaneous description of soft and semi-hard data   \label{sec:3.4}}

Having tuned the parameters to describe the available high-energy `soft' data ($\sigma_{\rm tot},~d\sigma_{\rm el}/dt,$ $d\sigma_{\rm SD}/dtdM^2,..$), we check whether the model can simultaneously describe semi-hard phenomena, such as the inclusive single-particle $p_t$ distribution, the gluon PDF and diffractive gluon PDF in the low $x$ domain.  Before presenting the detailed results, it is informative to first gain some insight into how the model offers the opportunity to simultaneously describe such diverse data.

\subsubsection{Choice between hypotheses (a) and (b) for multi-Pomeron couplings, $g^n_m$   \label{sec:pt}}
We saw, in subsection \ref{sec:proc}, that the amplitude, $F(y,\k,\b)$, is obtained from a perturbative QCD Leading Logarithmic (LL) formalism with an infrared cutoff, $q_0$. The low $k_t$ region, $k_t\gapproxeq q_0$, gives the 'soft' contribution, while at larger $k_t$ we have the conventional perturbative QCD result.

Note that in our approach the {\it same} parameter $\lambda$ simultaneously describes (i) the absorptive effects which suppress the low $k_t$
contributions, so pushing the majority of partons into a larger $k_t$ 
region, and (ii) proton dissociation into high-mass systems with $M^2 \gg m^2_N$ since the value of $\lambda$ determines the size of the triple- and multi-Pomeron contributions.
 Data corresponding to these processes, $d\sigma_{\rm SD}/dtdM^2$, are available
at Tevatron \cite{CDFhm,dino} and lower energies \cite{PhysRep}.
 As noted above, these are the data which fix the value of $\lambda$, or rather $f_{3p}$ defined below.
In this way we obtain a
{\it parameter-free} prediction for low $k_t$ suppression and thus for the  whole $\k$ behaviour (modulo simplifications used in the present model). 

Clearly, the $k_t$ dependence (the `additional' variable) along the ladder plays an important physical role. Note that now $\lambda$ cannot be considered just as a constant number. Its                           value depends on $k_t$.
It is reasonable
 to use the LL expression 
corresponding to the BFKL triple-Pomeron vertex
\be
\lambda=f_{3p}N_c\alpha_s(k_t)\pi^2\; ,
\label{eq:g3-bk}
\ee
where the constant $f_{3p}$, which allows for the possible renormalization caused by the next-to-leading corrections, may be regarded as the new parameter in  place of $\lambda$. Clearly we expect $f_{3p} \sim O(1)$.   $\lambda$ is dimensionless, since the absorption during the evolution is written in terms of the opacity $\Omega(y,\k,\b)$. The small probability of large $k_t$ partons rescattering is due to the decrease of $\Omega(\k)$ with increasing $k_t$.

Analogously, the starting distribution for the evolution of $F$ 
is written in terms of the LO (Born) cross section, which decreases as $1/k_t^2$, multiplied by the `input' parton density
\be
F(y_0,k_t^2,\b)~=~N'\beta(\b)N_c\alpha_s(k_t)/\pi k^2_t \ .
\label{eq:inputev}
\ee
 Here $N'$ is the normalization corresponding to the number of input gluons at $x=\exp(-y_0)$, and $\beta(\b)$  describes the parton distribution in the impact parameter plane. It is given by Fourier transform of the effective two-gluon form factor $\beta(t)$
\be
\beta(\b)=\frac{1}{4\pi^2}\int e^{i\vec{Q}_t\cdot \b} \beta(t) d^2 Q_t,
\label{eq:FT}
\ee
where $t=-Q_t^2$, and where we use the parametric form 
\be
\beta(t)=e^{c_2 t}/(1-t/c_1)^2.
\label{eq:c1c2}
\ee


Recall that, due to colour coherence, a parton with high $k_t$ in the ladder cannot be screened by one with lower $k_t$. The lower $k_t$ amplitude, with large wavelength, `sees' a point-like colourless object at high $k_t$, and does
not interact with it. Therefore the 'effective' value of the triple-Pomeron vertex, and the strength of suppression (due to the multi-Pomeron interactions) decreases with increasing $k_t$. 

The major contribution to the total and diffractive dissociation cross sections comes from the lowest $k_t$, close to the infrared cutoff $q_0$. In the case of $\sigma_{\rm tot}$ it is possible to compensate a larger value of $q_0$ by a larger input parton density $N'$. However the relative value of high-mass single dissociation,
$d\sigma_{\rm SD}/dM^2$  still decreases if the value of $q_0$ is increased.

Now, the multi-Pomeron couplings specified by hypothesis (a), given in subsection \ref{sec:hyp}, correspond to strong enhanced absorptive factors of the form $\exp(-\lambda\Omega)$. For this reason, we need to choose a rather low infrared cutoff $q_0=0.5$ GeV   
to obtain a satisfactory description of $\sigma_{\rm el}$, $d\sigma_{\rm el}/dt$, and high-mass dissociation, $d\sigma_{\rm SD}/dM^2$. 
Since the normalization is fixed by the soft 
cross sections at low $p_t\sim q_0$, and since all the higher $p_t$
contributions are suppressed by the ratio $q_0^2/p_t^2$, we fail to reproduce the inclusive single-particle spectra, $d\sigma/dydp^2_t$, in the larger
$p_t$ domain\footnote{The formula for the calculation of the $p_t$ distribution are given in (\ref{eq:pt})
of Appendix \ref{sec:B3}}. Indeed, the model underestimates the inclusive cross sections measured
at CERN and the Tevatron by about factor of 3-8 for $p_t>4$ GeV.
Simultaneously, the diffractive gluon PDF predicted by the model turns out to be a factor 3-5 smaller than that determined by the diffractive deep inelastic data measured at HERA.
These are rather general features of hypothesis (a). They do not change too much under variation of the parameters of the model, including the structure of the Good-Walker eigenstates
\cite{GW} used to reproduce low-mass proton dissociation.
Therefore below we will concentrate on the case (b).

\begin{figure}
\begin{center}
\includegraphics[height=8cm]{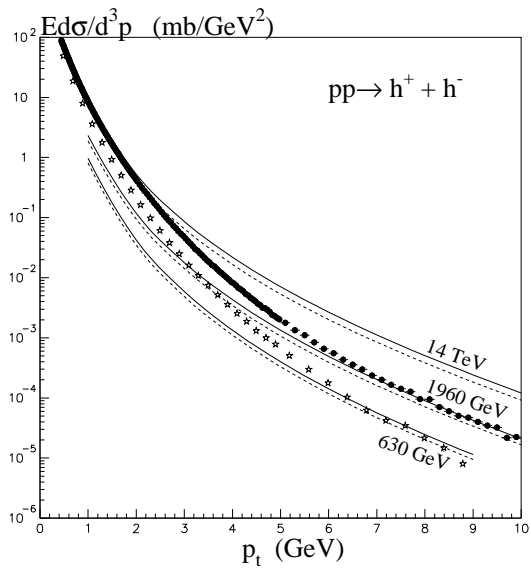}
\caption{\sf The description of the Tevatron data \cite{data} for the single-particle $p_t$ distribution, and the prediction for the LHC energy of 14 TeV. The continuous and dashed curves correspond to choices (ii) and (i), respectively of the diffractive eigenstates, see Section \ref{sec:4.1}.}
\label{fig:pT}
\end{center}
\end{figure}

With the multi-Pomeron couplings specified by hypothesis (b), the suppression of the low $p_t$ domain is weaker.
Here, a good description of `soft' cross sections, $\sigma_{\rm tot},\, 
d\sigma_{\rm el}/dt,\,  d\sigma_{\rm SD}/dM^2$, is obtained with the infrared cutoff chosen\footnote{This value of the infrared cutoff was used in a series of Durham 
papers for exclusive Higgs boson production and other Central Exclusive Diffractive processes \cite{KMR}.} to be $q_0=0.85$ GeV.  This increase of the cutoff from 0.5 to 0.85 GeV also makes the description of the single-particle inclusive cross section much better, see Fig~\ref{fig:pT}.
Now the predicted values $d\sigma/dydp^2_t$ are consistent with the present collider data for $p_t \gapproxeq 5$ GeV. At lower $p_t$ the naive prediction, based on a 
simple fragmentation of the gluon minijets, without a more precise treatment of hadronization, still underestimates the single-particle inclusive cross section. However, the problem of the deficit of low $p_t$ hadrons may be solved by accounting for the particles produced
via hadronization (that is the breaking) of the colour (LUND) strings. The role of this effect cannot be evaluated analytically. Here, we need a Monte Carlo.

\subsubsection{The gluon PDFs: $g(x,\mu^2)$ and $g^D(x_P,z,\mu^2)$    \label{sub:gPDF}}
Another check of how well our model can be matched to perturbative QCD, are the predictions it gives for the gluon PDFs. The `effective' gluon distribution may be calculated assuming that each $\k$ component corresponds to a ladder with a gluon of transverse momentum $\k$. Such a ladder gives us the gluon distribution $F(y,\k,\b)$ {\it unintegrated} over $\k$ and $\b$. 
Indeed, in terms of the unintegrated gluon distribution $f_g(x,\k^2,\mu^2)$, the integrated distribution is
\be
xg(x,\mu^2)~=~\int^{\mu^2}\frac{d\k^2}{\k^2}f_g(x,\k^2,\mu^2)\ ,
\label{eq:gluPDF}
\ee 
where the gluon density, $f_g$, unintegrated over $\k$ collects the gluons from the whole impact parameter $\b$ plane. That is
\be
\frac{f_g(x,\k^2,\mu^2)}{\k^2}~=~\int d^2\b F(x,\k^2,\b,\mu^2)
\ee
and
\be
xg(x,\mu^2)~=~\int d^2 \b\int^{\mu^2}d\k^2F(y,\k,\b),
\label{eq:glut}
\ee 
where $y={\rm ln}1/x$.
 However, care needs to be taken with this identification.
In order to compare with the gluons obtained from global parton analyses, that is gluons obtained from fits to DIS and related `hard'  data\footnote{We emphasize that the parameters of our model were tuned to `soft' $pp$ and $p\bar{p}$ data, not to DIS data.}, we have to calculate the unintegrated distribution $f_g$
 in the target proton neglecting the absorption caused by the beam opacity
$\Omega_i$. Thus we must use an equation of the form of (\ref{e15}) but with $\Omega_i=0$, such that there is no absorption caused by the beam proton. We  denote the target opacity obtained in this way by $\underline{\Omega}$, and the unintegrated gluon density by  $\underline{F}$. Hence the integrated gluon density is actually
\be
xg(x,\mu^2)~=~\sum_k|a_k|^2\int d^2 \b\int^{\mu^2}d\k^2 \underline{F}(y,\k,\b),
\label{eq:glutt}
\ee 
where the sum over the Good-Walker eigenstates $k$ in the incoming proton wave function $|p\rangle=\Sigma a_k|k\rangle$ is included\footnote{All the coefficients in the decomposition $|p\rangle =\sum a_k|\phi_k\rangle$
are taken to be $a_k=1/\sqrt 2$ (with $k=1,2$).}.

The so-called diffractive gluon distribution, $g^D(x_P,z,\mu^2)$, can also be predicted. In order to calculate 
$g^D(x_P,z,\mu^2)$ we have to replace the full opacity in (\ref{eq:glutt})
by the opacity corresponding to events with a rapidity gap between the target $k$ and an intermediate parton $c$ placed at $y=y_P=\ln(1/x_P)$, that is, to events with elastic $c-k$ scattering.  For the favoured hypothesis, (\ref{eq:(b)}), of the behaviour of the multi-Pomeron couplings, $g^n_m$, the starting value of $F_k^D$ at $y=y_P$ is given by
\be
F^D_k(y_P,\vec{k}_t,\vec b)=\frac{\Delta}{\lambda}
\left(1-e^{-\lambda\underline\Omega_k(y_P,\k,\b)/2}\right)^2\, ,
\ee
while for hypothesis (\ref{eq:(a)})
\be
F^D_k(y_P,\vec{k}_t,\vec b)=\Delta e^{-\lambda\underline\Omega_k(y_P,\k,\b)/2}
\left(1-e^{-\lambda\underline\Omega_k(y_P,\k,\b)/2}\right)
\underline F_k(y_P,\k,\b)/2\, .
\label{eq:gdiff}
\ee
After this, the diffractive density $F^D$ is evolved according to (\ref{e15}) with $\Omega_i=0$.

The predictions of the gluon PDFs using the model, with the favoured hypothesis (b) for the multi-Pomeron couplings, are also encouraging for the attempt to obtain a simultaneous description of soft and semi-hard phenomena. 
The diffractive gluons, $g^D(x_P,z,\mu^2)$, predicted by the model are in broad agreement with those obtained from the NLO analyses of the HERA diffractive deep inelastic data. Since the model accounts for enhanced absorptive corrections (which is a non-linear effect), the diffractive gluons grow with the scale $\mu^2$ a bit more slowly than those obtained by a NLO analysis based on linear DGLAP evolution.  The same is true for the inclusive integrated gluons $g(x,\mu^2)$. Actually the inclusive gluon distribution predicted by the model turns out to be between the values obtained from the LO and NLO global parton analyses of deep inelastic and related hard scattering data. As discussed above, it is surprising that the spectra of secondaries are better described by the Monte Carlo models which use LO and not NLO gluons. The detailed comparison of the predictions of the gluon PDFs with the previously known values is given in Tables \ref{tab:A3} and \ref{tab:A4} of the next Section. 

To conclude this Section, we emphasize that both the global inclusive and the diffractive gluon distributions were {\it not fitted},  but are {\it calculated} in the model, based on the values of the parameters which provided the best description of the purely soft cross sections ($\sigma_{\rm tot},\, d\sigma_{\rm el}/dt,\,  d\sigma_{\rm SD}/dM^2$). That is, within the model, the same values of the parameters control  the $x$- and scale-behaviour of the gluon PDFs, as well as the suppression of the inclusive single-particle  distribution at low $p_t$, as discussed in subsection \ref{sec:pt}.

\section{Results of the fit to soft data and predictions for semi-hard phenomena}

In this section we present the description of the `soft' data obtained by tuning the parameters of the model, as well as the detailed predictions for the gluon PDFs. The soft data on their own do not definitively distinguish between hypotheses (a) and (b) of Section \ref{sec:hyp} for the form of the multi-Pomeron couplings. However, as explained in the previous Subsection \ref{sec:3.4}, the two sets of predictions for the semi-hard phenomena strongly favour hypothesis (b), with the infrared cutoff $q_0=0.85$ GeV. We therefore present results for just this hypothesis. First, we give the values of the parameters that we obtain.

\subsection{The parameters and their values   \label{sec:4.1}}
The list of the, remarkably few, parameters of the model was given in Section \ref{sec:para}. Moreover, the parameters are physically motivated, and for many we have a good idea what their value should be. We have already defined $\Delta$ in (\ref{eq:Delta}), $N'$ in (\ref{eq:inputev}), $c_1$ and $c_2$ in (\ref{eq:c1c2}), $\lambda$ (which was replaced by $f_{3p}$) in (\ref{eq:g3-bk}), $q_0$ and $y_0$ at the end of subsection \ref{sec:proc}.  The diffusion parameter $d$ will be specified in (\ref{eq:K}) below.  


Now let us define the parameters $\gamma_1$ and $\gamma_2$, which are needed to account for low-mass proton dissociation.
The incoming proton wave function is written as a sum of two components\footnote{A two-channel eikonal was shown to suffice, see \cite{KMRns}.}: the Good-Walker diffractive eigenstates, $i=1,2$. These states are preserved (and not mixed) by an interaction with the Pomeron. That is, they undergo elastic-like scattering. The ratio of the
Pomeron couplings to the first and the second state was chosen to reproduce the observed value of the low-mass single dissociation cross section\footnote{Here, $\sigma_{\rm SD}^{{\rm low} M}$ corresponds to data with the mass of the dissociating system $M<2.5$ GeV, which approximately equates to $y \lapproxeq 1.5$. These data are, to our knowledge, the only available measurement of $\sigma_{\rm SD}^{{\rm low} M}$, and even this has large uncertainties. Thus the identification of the diffractive eigenstates is far from unique.} $\sigma_{\rm SD}^{{\rm low} M}=2$ mb at the CERN-ISR energy of $\sqrt{s}=31$ GeV \cite{CERN-ISR}. However, still some freedom remains. In general, each eigenstate may have its own form factor, which specifies the distribution of the partons in $b$ space. We assume that the distribution has the same form 
for each state, except for the radius of the state, $R_i$. We consider two 
 possibilities: 

(i)  $R^2_i\propto \sigma_i$, ~~that is, the same 
parton density at the origin ($b=0$) of each state, 

(ii) 
$R_i\propto \sigma_i$, ~~which corresponds to a BFKL-like coupling
 $\beta_i\propto 1/k_i$. 

In both cases, after tuning  
the parameters, we obtain qualitatively (and mainly quantitatively) the same results. The main difference is that the model with the BFKL-type couplings of (ii) predicts the diffractive PDF of the gluon to be closer to the conventional values obtained from the diffractive HERA data. Although (ii) is the favoured choice we present the results for both possibilities for the $\gamma_i$'s. The values of the parameters are listed in Table \ref{tab:A1}. 
\begin{table}[htb]
\begin{center}
\begin{tabular}{|l|c|c|}\hline
 &   (i) &  (ii)  \\ \hline

 $\Delta$  &   0.31   &     0.33    \\
 $N'$    &  6.8   &     6.85       \\
 $c_1$   &  0.95  &     3        \\
 $c_2$   &  0.1  &     0.9        \\
$f_{3p}$   &  0.6  &     1        \\
$\gamma_1$   &  1.57  &     1.5        \\
$\gamma_2$   &  0.43  &     0.5      \\
\hline

\end{tabular}
\end{center}
\caption{\sf The values of the parameters for assumptions (i) and (ii) for the diffractive eigenstates. The other parameter values are $d^2=2$,  $y_0=1.5$ and $q_0=0.85$ GeV; also GeV units are used for $c_1$ and $c_2$.}
\label{tab:A1}
\end{table}

\subsection{The description of the `soft' data}
The values of the total, elastic and  single proton dissociation cross sections are given in Table \ref{tab:A2} for collider energies $\sqrt{s}=1.8,~7,~14, ~100$ TeV.  The detailed formulae used to calculate the cross sections are given in Appendix B.
\begin{table}[htb]
\begin{center}
\begin{tabular}{|c|c|c|c|c|c|c|}\hline
energy &   $\sigma_{\rm tot}$ &  $\sigma_{\rm el}$ &    $\sigma_{\rm SD}^{{\rm low}M}$ &  $\sigma_{\rm SD}^{{\rm high}M}$  &   $\sigma_{\rm SD}^{\rm tot}$  &  $\sigma_{\rm DD}^{{\rm low}M}$ \\ \hline

 1.8  &   72.8/72.5   &     16.3/16.8  &       4.4/5.2      &   7.0/7.8  &     11.4/13.0   &  0.3/0.4         \\
 7   &   89.0/86.8    &    21.9/21.6    &     5.5/6.7   &      9.9/10.2  &     15.4/16.9  &   0.5/0.7  \\
14    &  98.3/94.6   &     25.1/24.2     &    6.1/7.5   &    11.5/11.3  &     17.6/18.8    &  0.6/0.9  \\
 100   &  127.1/117.4  &     35.2/31.8  &       8.0/9.9   &     16.7/14.4  &     24.7/24.3  &  0.9/1.6   \\
 \hline

\end{tabular}
\end{center}
\caption{\sf Cross sections (in mb) versus collider energy (in TeV). The first number corresponds to choice (i), and the second number corresponds to choice (ii) of the diffractive eigenstates. $\sigma_{\rm SD}$ denotes the sum of single-dissociation of the beam and the target.}
\label{tab:A2}
\end{table}

\begin{figure} [t] 
\begin{center}
\includegraphics[height=12cm]{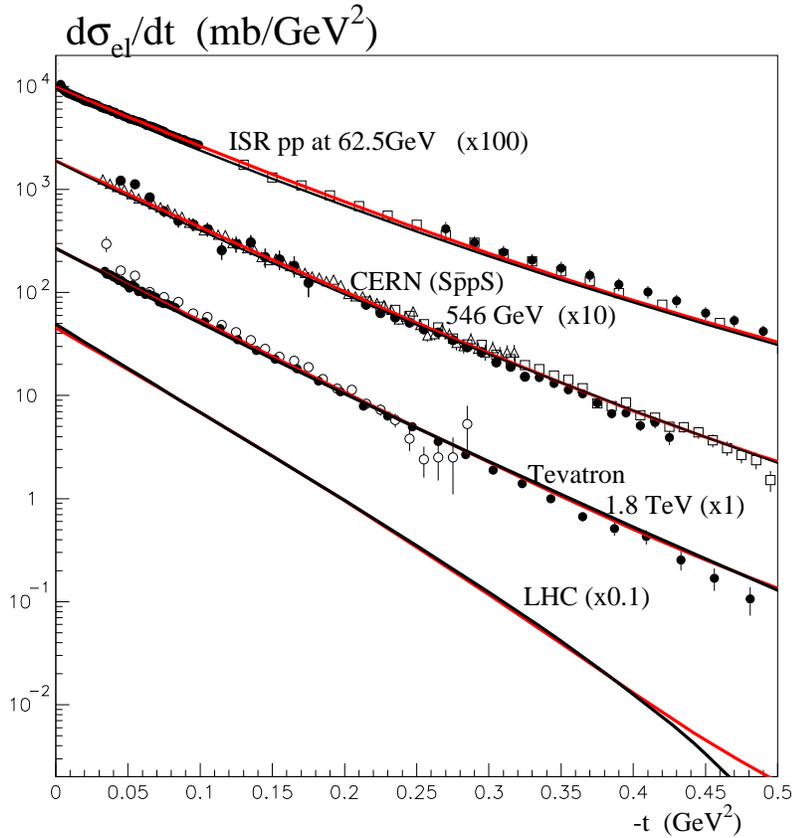}
\caption[*]{\sf The $t$ dependence of the elastic $pp$ cross section, and the prediction for 14 TeV. The bolder and fainter (red) curves correspond to choices (i) and (ii) of the diffractive eigenstates respectively. The references for the data are the same as those given in \cite{KMRsoft}.} 
\label{fig:dsdt}
\end{center}
\end{figure}

We show in Fig.~\ref{fig:dsdt} the quality of the description of the data for the {\it elastic differential} cross section.
In the tuning of the parameters to describe these data, we find that the slope of the  $t$-distribution at CERN-ISR energies is a bit too steep, and that it is a bit too flat at the Tevatron energy. This indicates the need to add a small non-zero slope, $\alpha'$, of the bare Pomeron trajectory, which was neglected in the present computations\footnote{In particular, a small non-zero contribution to $\alpha'$ is generated by the pion-loop insertion into the Pomeron trajectory \cite{angr,KMRsoft}.}.
  We also show 
 the prediction for differential elastic cross section at
an LHC energy of $\sqrt{s}=14$ TeV.

\begin{figure} 
\begin{center}
\includegraphics[height=16cm]{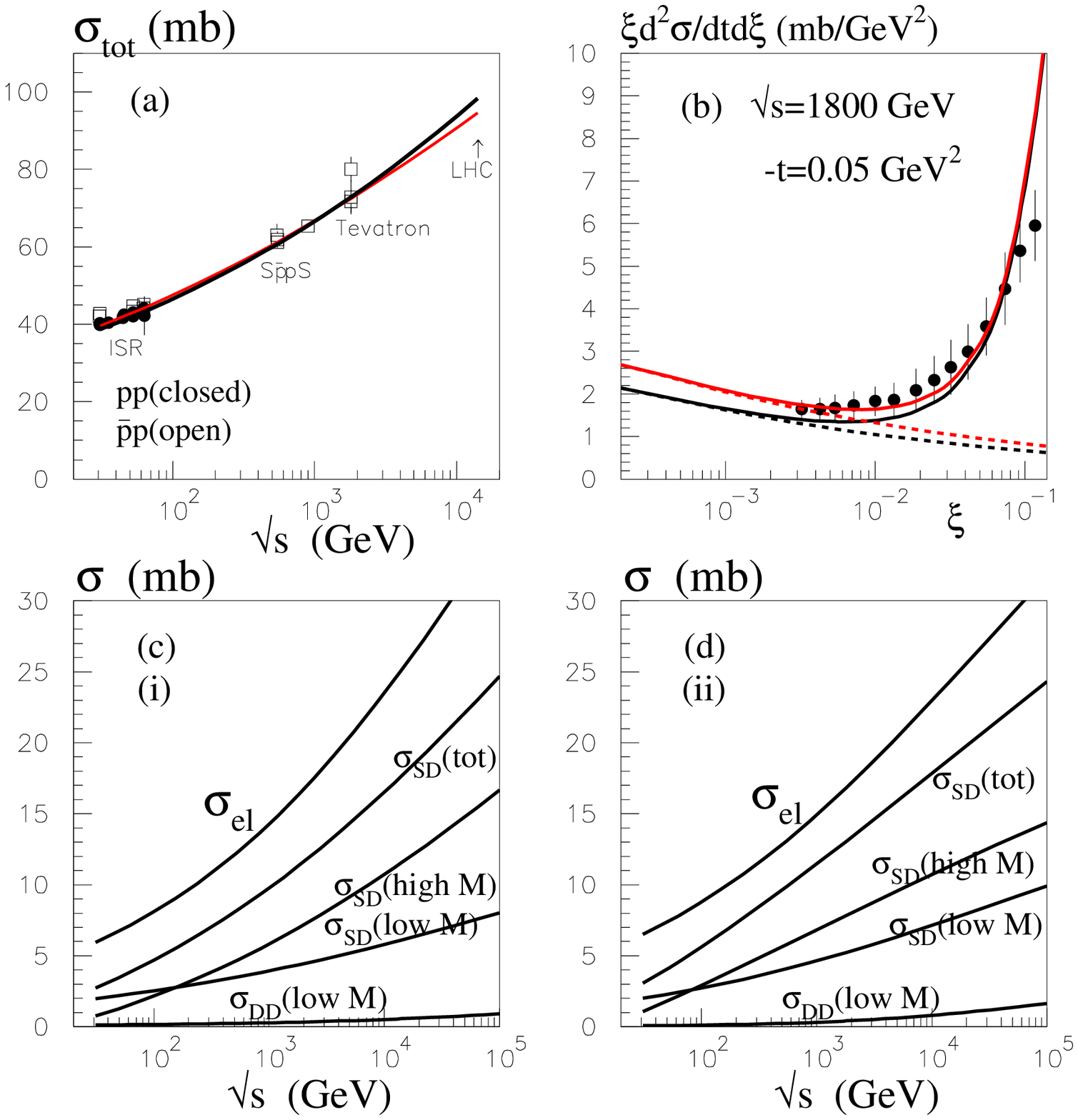}
\caption[*]{\sf The description of the data (a) for the total $pp(p\bar{p})$ cross section, and (b) for single dissociation. The bolder and fainter (red) curves correspond to choices (i) and (ii) of the diffractive eigenstates respectively. The dashed lines in plot (b) correspond to the result if the secondary Reggeon contributions were to be neglected. Plots (c),(d) show the energy dependence of elastic and diffractive cross sections for choices (i),(ii). $\sigma_{\rm SD}$ denotes the sum of single-dissociation of the beam and the target.}
\label{fig:sum4}
\end{center}
\end{figure}

The description of the total cross section data is shown in Fig.~\ref{fig:sum4}(a). The screening corrections, which arise from the `enhanced' multi-Pomeron
  diagrams, that is, from the high-mass dissociation, slow down the growth of
  the cross section with energy.  Thus, the model predicts a  relatively low
total
cross section at the LHC energy of 14 TeV:
\be
\sigma_{\rm tot}\simeq 95-100~ {\rm mb}.
\ee 

\begin{figure} 
\begin{center}
\includegraphics[height=14cm]{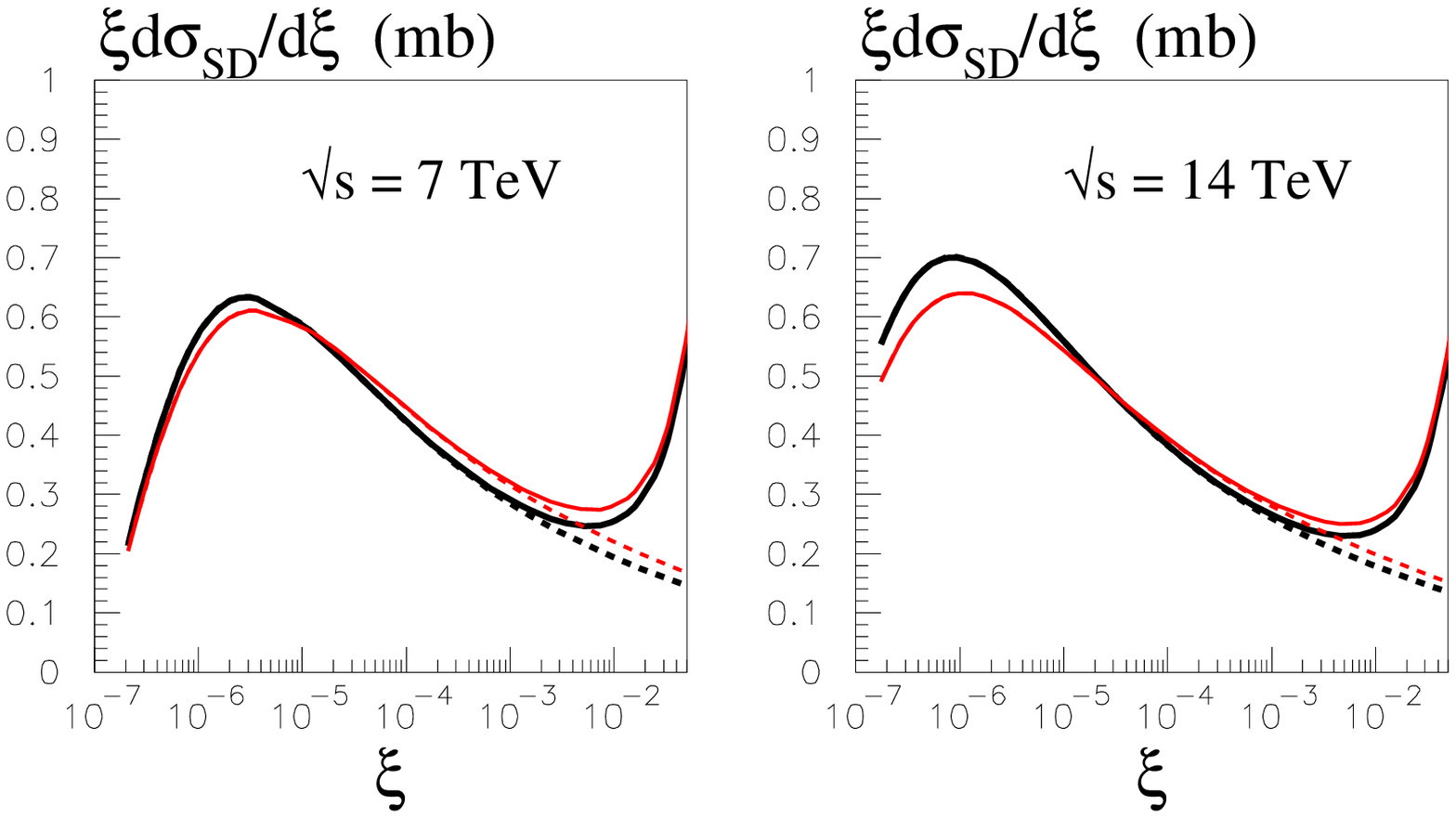}
\vspace{-7cm}

\caption[*]{\sf Cross sections for single proton dissociation, $\xi d\sigma_{\rm SD}/d\xi$, integrated over $t$, for two LHC energies. The bolder and fainter (red) curves correspond to choices (i) and (ii) of the diffractive eigenstates respectively.  The dashed curves are obtained if the contributions of the secondary Reggeons are omitted. The secondary Reggeon contributions are computed as in \cite{KMRnns1}. These contributions are not, of course, included in the values of $\sigma_{\rm SD}$ listed in Table \ref{tab:A2}; formally, they are not of diffractive origin.}
\label{fig:SD}
\end{center}
\end{figure}

 The energy behaviour of the elastic and diffractive  cross sections are shown in Table 2 and Fig.~\ref{fig:sum4}(c,d), where the high-mass single dissociation cross sections include all events with $M^2/s<0.05$.
The dependence of the cross section for high-mass dissociation, $\xi d^2\sigma/dtd\xi$, on $\xi=M^2/s$ is compared with the Tevatron CDF data \cite{CDFhm,dino} in Fig.~\ref{fig:sum4}(b). Recall that in our model we have not included the secondary Reggeon contributions, and the corresponding results are shown by the dashed lines. The continuous lines include the secondary Reggeon contributions computed as in \cite{KMRnns1}.

\subsection{Proton single dissociation at the LHC}
The energy behaviour of diffractive cross sections is shown in Fig. \ref{fig:sum4}(c,d) for the two choices of radii, (i,ii), of the Good-Walker diffractive eigenstates. Due to a small probability, $S^2$, that the rapidity gap survives rescattering in the centre of the disk, the cross section of diffractive dissociation comes mainly from large impact parameters, $b$, that is from the periphery. Now, in the favoured model (ii),  the
component with a large inelastic cross section has a much larger
radius and a smaller parton opacity. Therefore, model (ii) predicts noticeably larger
diffractive dissociation.

At this point, it is worth emphasizing again that the available soft data at pre-LHC energies are not sufficient to fix the details, and the precise values of the parameters,
of the model. Another example of this, is the Ostapchenko model \cite{ost}, which is rather close ideologically to our approach\footnote{The Ostapchenko model is discussed in Section \ref{sec:ost}.}.  It also fits the pre-LHC data. Nevertheless, with set (A), Ostapchenko predicts that $\sigma_{\rm tot}$ is about 30 mb larger, and  $\sigma^{{\rm high}M}_{\rm SD}$ is more than twice  smaller, than our expectations  at $\sqrt {s}=14$ TeV.

As mentioned in Subsection \ref{sec:4.1}, the identification of the diffractive (Good-Walker) eigenstates is a major uncertainty, since we have little experimental information on low-mass dissociation. Measurements of low-mass proton dissociation at the LHC can possibly be
obtained using the Zero Degree Calorimeter and the Forward Shower
Counters \cite{fsc}.  This would shed light on the structure of the diffractive
eigenstates and help to better constrain the model. 

The expected cross sections of high-mass single diffractive dissociation, integrated over the transverse component of the momentum transferred to the proton (that is, integrated over $t$), are shown in Fig. \ref{fig:SD} at the LHC energies of 7 and 14 TeV. Here we have plotted the results down to very small $\xi$; $\xi \sim 10^{-7}$. It is clear that measurements in the region $\xi \gapproxeq 10^{-6}$ would be very valuable. 
At present such small  $\xi$ values
 cannot be measured by the forward proton detectors,
but we may expect that such events with very large rapidity gaps 
 can be
 selected, using  forward calorimeters and, hopefully, the  Forward
 Shower Counters \cite{fsc}.
 For $\sqrt s=14$ TeV,  $\xi=10^{-6}$ corresponds to the production of a diffractive system of the mass $M=14$ GeV. The position of the edge
of the large rapidity gap, $y\simeq 4$, may be observed in the calorimeter.

An interesting, but a little unusual, feature of the plots \ref{fig:SD}(a) and (b) are the maxima at $\xi\sim 3\times 10^{-5}$ and $\xi\sim  10^{-6}$, that is, at $M\sim $ 12 and 14 GeV, for the two LHC energies respectively.
In the simplified triple-Pomeron approach one expects a growth of $M^2d\sigma/dM^2\propto 1/(M^2)^\Delta$ with $M$ decreasing; we assume here a Pomeron intercept $\alpha(0)=1+\Delta$. Such behaviour (with the intercept `renormalized' by more complicated, enhanced diagrams) is, indeed, observed in our model in the $\xi = 10^{-5}- 10^{-3}$ interval. However, at very small $\xi$, the probability of dissociation starts to decrease with decreasing $\xi$. The explanation is that the cross section coming from low $k_t$ components is strongly suppressed by enhanced absorptive effects. Thus most of the high-mass dissociation comes from relatively high $k_t$. On the other hand, at small $M$, the rapidity ($\ln 1/x$) interval and available phase space is insufficiently large to generate high $k_t$ partons starting from the evolution of `soft' partons in the proton wave function. This explanation is verified by using the same model, but without the $k_t$ dependence\footnote{Allowing only a very small interval of $k_t$ variation.}. Then the maximum disappears and $\xi d\sigma/d\xi$ continues to increase with decreasing $\xi$. The growth of the cross section for high-mass dissociation with increasing $M$, 
at a relatively low $M^2$, is a qualitatively new feature of a model which explicitly accounts for the $k_t$ dependence of the absorptive effects during the evolution.

\begin{table} 
\begin{center}
\begin{tabular}{|c|c|c|c|c|}\hline
 & $\mu^2=4$ & $\mu^2=16$ &$\mu^2=64$ &$\mu^2=256$ \\ \hline
$x=10^{-2}$  & 5.2/5.3 & 9.8/10.1 & 15.3/15.9 & 22.1/23.2   \\
 & (3.3-5) & (5.2-6) & (6.3-7) & (7.1-7.5)  \\  \hline
$x=10^{-3}$  & 6.7/6.9 & 14.1/14.7 & 24.4/26.0 & 39.0/42.0   \\
 & (3.8-9) & (9.6-16) & (15-22) & (19.6-27)  \\ \hline
$x=10^{-4}$  & 9.1/9.3 & 21.0/22.1 & 39.9/43.2 & 68.8/75.8   \\
 & (4.2-16.5) & (16.1-36) & (29.8-56) & (44.4-75)  \\  \hline
\end{tabular}
\end{center}
\caption{\sf The values of the integrated gluon distribution, $xg(x,\mu^2)$ with $\mu^2$ in units of ${\rm GeV}^2$, computed using the soft model, as given by eq.(\ref{eq:glutt}). As in Table \ref{tab:A2}, the two numbers correspond to choices (i) and (ii) for the diffractive eigenstates. For comparison the numbers in brackets correspond to the integrated gluon distribution determined in the MSTW NLO (first) and LO (second) global parton analysis \cite{MSTW}.}
\label{tab:A3}
\end{table} 
\begin{table} 
\begin{center}
\begin{tabular}{|c|c|c|c|c|}\hline
 & $\mu^2=4$ & $\mu^2=16$ &$\mu^2=64$ &$\mu^2=256$ \\ \hline
$x_P=0.005$  & 0.16/0.26 & 0.20/0.35 & 0.24/0.42 & 0.29/0.51   \\
 $z=0.2$ & (0.23) & (0.37) & (0.42) & (0.43)  \\  \hline
$x_P=0.005$  & 0.18/0.30 & 0.26/0.44 & 0.35/0.62 & 0.48/0.85   \\
 $z=0.05$ & (0.24) & (0.53) & (0.72) & (0.85)  \\ \hline
$x_P=0.001$  & 0.24/0.40 & 0.33/0.56 & 0.40/0.70 & 0.49/0.86   \\
 $z=0.2$ & (0.31) & (0.54) & (0.65) & (0.68)  \\  \hline
$x_P=0.001$  & 0.27/0.44 & 0.40/0.69 & 0.56/0.96 & 0.75/1.32   \\
$z=0.05$ & (0.32) & (0.76) & (1.09) & (1.30)  \\  \hline
\end{tabular}
\end{center}
\caption{\sf The values of the diffractive gluon distribution, $x_Pzg(x_P,z,\mu^2)$ with $\mu^2$ in units of ${\rm GeV}^2$, computed using the soft model, as given by eqs.(\ref{eq:glutt},\ref{eq:gdiff}). As in Table \ref{tab:A2}, the two numbers correspond to choices (i) and (ii) for the diffractive eigenstates. For comparison the numbers in brackets correspond to the diffractive gluon distribution determined from the MRW analysis of the diffractive HERA data \cite{MRW}.}
\label{tab:A4}
\end{table}

\subsection{The gluon PDFs: $g(x,\mu^2)$ and $g^D(x_P,z,\mu^2)$}

In Tables \ref{tab:A3} and \ref{tab:A4} we present the predictions for inclusive and diffractive gluon densities respectively, obtained from the parameter values listed in Table \ref{tab:A1}. A discussion of the comparison of the predicted values of the gluon PDFs with the conventional determinations was given in subsection \ref{sub:gPDF}

\subsection{Rapidity gap survival probability}
The above model also allows us to calculate the probability that the rapidity gaps in diffractive processes survive both eikonal and enhanced rescattering.
Recall that the absorption caused by rescattering of intermediate partons is included as the factors (\ref{sap-b}), or (\ref{sap-a}), in equations (\ref{e15},\ref{e16}) for the opacities, depending on whether hypothesis (b), or (a), for $g^n_m$ is adopted. 
 So we have the possibility to quantify the role of the 
  suppression of processes with rapidity gaps due to enhanced rescattering, by solving the evolution 
 equations with and without the enhanced absorption factors.
We denote the `target' $k$ opacity, calculated without the
absorptive factor  from the beam $i$ side, as 
$\underline{\Omega}_k$; and the corresponding 
matrix element as 
$\underline{\cal M}$. Further, if we denote the matrix element which includes the 
absorptive  factors as
${\cal M}^{\rm enh}$, then we obtain \cite{KMRnns2}
\be
S^2_{\rm enh}~=~\frac{\left| {~\cal M}^{\rm enh} \right|^2}{\left| ~\underline{\cal M}~ \right|^2}.
\ee 
Note that now the value of $S^2_{\rm enh}$ can be calculated for any fixed
 virtuality (transverse momenta) of the partons which initiate the 
hard subprocess. Then it can be included {\it inside} the $k_t$ integral 
for the hard subprocess amplitude. In this way, we can account precisely for the 
`soft-hard factorization breaking' in central exclusive production
and other processes.
\begin{table}[htb]
\begin{center}
\begin{tabular}{|r|c|c|}\hline
 energy &   (i) &  (ii)  \\ \hline

7 TeV  &   0.013   &     0.024    \\
14 TeV    &  0.008   &     0.015       \\
 \hline

\end{tabular}
\end{center}
\caption{\sf The effective (corresponding to the slope $B=4$  GeV$^{-2}$) survival factor, $\langle S^2 \rangle$, of the rapidity gaps for the exclusive production, $pp \to p+H+p$, of a Standard Model Higgs boson of mass 120 GeV at two LHC energies.  Both eikonal and enhanced rescattering are taken into account.}
\label{tab:A2a}
\end{table}

Indeed, our model allows the calculation of the rapidity gap survival factor, $S^2(\b)$, for any diffractive process as a function of the impact parameter $\b$. A topical, and important, example is the
 central exclusive production of a Higgs boson via the process $pp \to p+H+p$, where the $`+'$ signs denote large rapidity gaps.  The overall survival factors $\langle S^2 \rangle$, are given in Table \ref{tab:A2a} at two LHC energies , for two choices, (i) and (ii), of the diffractive eigenstates; here the survival factors take into account both eikonal and enhanced rescattering. Recall that (ii) is the favoured choice. The values are in agreement with our previous determination at $\sqrt{s}$=14 TeV \cite{KMRnns2}
\be
\langle S^2 \rangle~=~0.015^{+0.01}_{-0.005}.
\ee
Note that, in order to compare with other results\footnote{Note that the very recent value of the survival factor obtained by the Tel-Aviv group \cite{GLM11} is an order-of-magnitude greater than their previous determination \cite{GLMM}, and more in agreement with our value.}, we have presented the effective values of 
$\langle S^2 \rangle$ which correspond to
 the $t$-slope $B=4~\GeV^{-2}$ of the {\it bare} hard cross section, that is, before accounting for absorptive effects. The value of $\langle S^2 \rangle$ increases with the slope $B$, but the ratio $\langle S^2 \rangle/B^2$ is approximately constant for reasonable variations of $B$; again more discussion is given in \cite{KMRnns2}.

\section{Comparison with the other approaches  \label{sec:ost}}

At the moment, the only other approach which includes a complete set of the 
multi-Pomeron vertices is the model presented by Ostapchenko \cite{ost}. A comparison with our model is   
instructive. As in our case, the Ostapchenko model uses a 2-channel (eikonal) 
Good-Walker formalism to account for low-mass diffractive dissociation, 
and the approach takes some account of the internal transverse-momentum structure of the Pomeron. 

First, recall, that in perturbative QCD, the BFKL Pomeron is not a pole, but a cut, in the complex angular momentum plane. This reflects the non-trivial structure of the BFKL vacuum singularity. Such a Pomeron may be treated as a series of components of different transverse size. Ostapchenko models this feature by simply introducing two different poles corresponding to small- and large-size Pomerons, whereas here we allow for the explicit $k_t$ dependence of the wave function corresponding to the $t$-channel Pomeron state; that is, we allow for the BFKL cut. In other words, the two poles of the Ostapchenko model may be considered as a way to mimic the BFKL cut, while we study the full $k_t$ dependence. Moreover in \cite{ost}, the two poles were combined into a single propagator. As a result, one cannot consider the effect of a single Pomeron pole, with the contribution of the other component decoupled. This is a 
 shortcoming of the model. In particular, in multi-Pomeron diagrams,
there is screening of the small-size Pomeron by the large-size Pomeron, while in perturbative QCD this effect is {\it absent}. To be more precise, at leading order, the
triple-Pomeron vertex contains (after the azimuthal integration) a $\Theta$-function (see \cite{theta}), which reflects the fact that a long-wavelength gluon (from the large-size Pomeron) interacts {\it coherently} with the small-size dipole formed from a pair of $t$-channel gluons in the ladder which represents the small-size BFKL Pomeron.  Therefore the coupling is proportional to the {\it whole} colour charge of this small-size dipole. Since the Pomeron as a whole is colourless, the vertex {\it vanishes}.

Contrary to our previous soft models \cite{KMRnns1,KMRsoft}, we have seen above that the introduction of constraints from semi-hard phenomena 
now favours multi-Pomeron couplings $g^n_m$ given by hypothesis (b) of (\ref{eq:(b)}), which, for the convenience of this discussion, we write as 
\be
g^n_m~=~ r_{3P}(g_N\lambda)^{n+m-3},
\label{eq:gnmo}
\ee
with $r_{3P}=g_N\lambda\Delta$.
The advantage of form (\ref{eq:(b)},\ref{eq:gnmo}) is the possibility to use the simplest generalisation of the AGK cutting rules \cite{AGK}. These rules determine the relative sizes of the contributions of the processes which result from the different cuts of the exchanged Pomerons (in a multi-Pomeron diagram) in terms of combinatorial factors only.

This same form (\ref{eq:gnmo}) was used in the Ostapchenko and  KPT~\cite{KPT}  models.
However Ostapchenko uses two parameters: $r_{3P}$ for the triple-Pomeron vertex and $\lambda$ to allow for the other multi-Pomeron couplings, whereas we consider the so-called `critical Pomeron' regime in which the value of $r_{3P}=g_N\lambda\Delta$ is fixed.

An attractive feature of our critical Pomeron approach is that the two-particle irreducible amplitude, $\Omega(\b)$ of (\ref{eq:Tb}), generated by the evolution equations (\ref{e15},\ref{e16}),
never grows as a power of energy, but instead
\be
\Omega(\b)\propto \ln s.
\ee
 On the contrary, if the general form (\ref{eq:gnmo}) with an {\it arbitrary} value of the coupling $r_{3P}$  were to be used, then a power-like asymptotic behaviour, $\Omega\propto s^{\alpha-1}$, is obtained with a renormalized intercept (see \cite{KPT} and Sect.4 of the first paper of \cite{ost})
\begin{equation}
\alpha^{\rm ren}_P=\alpha^{\rm bare}_P-r_{3P}/g_N\lambda
\label{eq:ren}
\end{equation}
Here $\alpha^{\rm bare}_P$ and $\alpha^{\rm ren}_P$ denote the Pomeron intercepts before and after including the enhanced multi-Pomeron diagrams.
If $\alpha^{\rm ren}_P<1$ then the cross section asymptotically decreases with energy. On the other hand, if the renormalized intercept $\alpha^{\rm ren}_P>1$ then unitarity is still not satisfied by the renormalized Pomeron exchange. In other words the enhanced contributions arising from  interactions  between the partons inside one parton cascade are not sufficient to provide  saturation and to stop 
the growth of parton densities. Unitarity is only satisfied after eikonalization of the final amplitude.

It is worth emphasizing that when parameters were tuned to describe the data, the Ostapchenko model resulted in an intercept $\alpha^{\rm ren}_P<1$ for the large-size Pomeron component, while the intercept of the small-size component became very close to one, for example, for set C it was found that $\alpha^{\rm ren}_P=1.03$ \cite{ost}. This would indicate that Nature  prefers the so-called `critical Pomeron' regime (with $\alpha^{\rm ren}_P=1$ resulting in a logarithmic energy behaviour of the irreducible amplitude), which is generated automatically by our form of the multi-Pomeron vertices.

Note also that models which allow only triple-Pomeron vertices,
such as that used by the Tel-Aviv group \cite{GLMM}, predict an asymptotically decreasing cross section.  The reason is that (\ref{eq:ren}) gives a negative $\alpha^{\rm ren}_P$ in the limit $\lambda\to 0$. However, the MPSI approximation used in \cite{GLMM} leads to a constant cross section, which indicates that the method is only applicable to a limited energy interval, see also \cite{ost}.
There should be the parameter which controls the validity of the  approach of Ref. \cite{GLMM}.
 Besides this the Tel-Aviv model accounts for only one component of the Pomeron and its $k_t$ structure was not considered at all.

\section{Implications for ultra-high-energy cosmic rays}

A crucial ingredient in the determination of the primary energy of ultra-high-energy cosmic rays, which generate extensive air showers in the Earth's atmosphere, is a detailed model of hadronic multiparticle production which is reliable up to very high centre-of-mass energies, more than an order-of-magnitude higher than that available at the LHC. This enables the development of the cascade to be modelled from the primary interaction high in the atmosphere to the observations on the Earth.

A recent paper \cite{UHECR} discusses the attributes of the Monte Carlos that have so far been used to generate these extensive air showers. The discussions in \cite{UHECR} show that none of these Monte Carlos are able to completely describe all the available measurements made at the high energy Tevatron and LHC colliders. The model we have described above has the potential to have all the required attributes, but it is still basically at the parton level and needs hadronization to be fully implemented. That is, a Monte Carlo realisation is necessary\footnote{The construction of such a Monte Carlo is, at present, underway \cite{Krauss}.}. However some comments can already be made.  Above, we have discussed the multi-Pomeron structure of the QGSJETII Monte Carlo of Ostapchenko \cite{ost}, which has been used to generate the extensive cosmic ray air showers.  We saw that set (A) of QGSJET predicts that $\sigma_{\rm tot}$ is about 30 mb larger, and,  even more importantly, that $\sigma^{{\rm high}M}_{\rm SD}$ is more than twice  smaller, than the  expectations of our model  at $\sqrt {s}=14$ TeV. 
It is clear that the difference between the predictions of the two models is large. Moreover, for the generation of cosmic ray air showers, proton diffractive dissociation is very important since it produces the leading hadrons which carry a large fraction, $x_L$, of the initial energy.
So confirmation, at the LHC, of the expectations for $\xi d\sigma/d\xi$, and especially the study of low-mass dissociation, would be extremely instructive.

\section{Conclusions}

 At first sight, it is surprising that the present `soft' LHC data are tolerably well described by  Monte Carlos based on  LO DGLAP evolution. 
 Actually, at such large collider energies 
 we expect BFKL-like evolution, where there is no strong $k_t$ ordering, to be more relevant.  Moreover, in DGLAP-based Monte Carlos,  the `soft' data are reproduced within a 'hard' approach with
 a rather high infrared cutoff $k_{\rm min}$ which increases with collider energy reaching a value $k_{\rm min}\simeq 3$ GeV at $\sqrt{s}$ = 7 TeV.

In this paper we explain this puzzle. We use a perturbative QCD framework based on Reggeon Field Theory. That is, an approach based on the Regge-like structure of the BFKL framework in which the symmetry of the kernel allows evolution with both strong $k_t$ and inverse $k_t$ ordering; in other words diffusion in ln$k_t$.
We demonstrate that the rescattering and absorption of the intermediate partons, described by the enhanced reggeon diagrams,
 strongly suppresses the evolution in the direction of low $k_t$,
 where the absorptive cross section $\sigma^{\rm abs}\propto 1/k^2_t$ is large. This acts as an effective infrared cutoff, $k_{\rm min}$, whose value grows with increasing energy. It thus explains the dominance of evolution in the `DGLAP
  direction', that is with $k_t$ increasing from the proton to the central rapidity region.
  Thus we have the opportunity to describe `hard' and `soft' processes in a unified way.
  
  The model is based on simplified LO perturbative QCD expressions,
  with `renormalization' parameters to account for the higher $\alpha_s$ order corrections, and embodies multi-Pomeron absorptive effects. It  can be extended into the low $k_t$ domain,
  and it allows a satisfactory simultaneous description of `soft'
  and `hard' phenomena. In particular, after the parameters are fixed to describe the behaviour and the absolute values of the available soft cross sections ($\sigma_{\rm tot}, d\sigma_{\rm el}/dt, d\sigma_{\rm SD}/dM^2$), the model generates
   quite reasonable low-$x$ gluon distributions (for  both the diffractive and inclusive processes) {\it without} fitting any additional parameters responsible for the `input' parton distributions. Recall that, on the contrary, in conventional global parton analyses the `input' distributions coming from the soft (confinement) region are completely unknown, and their forms are parametrised and fitted to describe the experimental results for DIS and related `hard' processes.
  
  Some predictions of the model for the LHC energies are given in Figs. \ref{fig:pT}$-$\ref{fig:SD} and Tables \ref{tab:A2}$-$\ref{tab:A2a}.

\section*{Appendices}

\appendix

\section{Evolution equations}
\begin{figure} 
\begin{center}
\includegraphics[height=6cm]{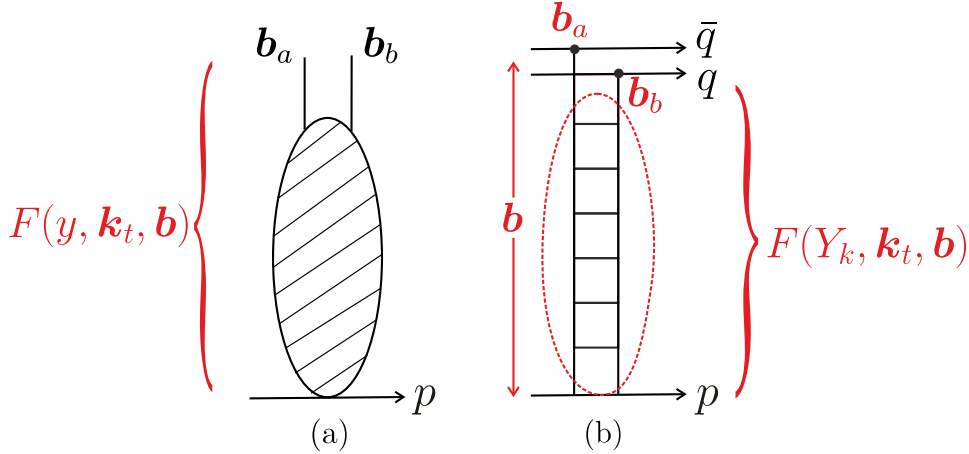}
\caption{\sf (a) The bare Pomeron exchange amplitude $F(y,\k,\b)$ in terms of colourless two-gluon exchange, where $\vec{b}=(\vec{b}_a + \vec{b}_b)/2$  and $\k$ is conjugate to $(\vec{b}_a - \vec{b}_b)$; (b) the evolution of the bare Pomeron amplitude in rapidity, for the example of a proton interacting with a colourless $q{\bar q}-$ dipole}
\label{fig:ladder2}
\end{center}
\end{figure}

Here we give more details of the model that we have outlined in Section \ref{sec:A3}. We start with the evolution equation, (\ref{eq:ev}), which generates the ladder structure of the bare Pomeron exchange amplitude $F(y,\k,\b)$, as shown in  
Fig.~\ref{fig:ladder2}(a). The amplitude depends on the positions, $\vec{b}_a$ and $\vec{b}_b$, of the gluons ($a,b$) in the impact parameter plane, and on the available rapidity interval, $y$. It is convenient to introduce the conjugate transverse momentum variable $\k$ to $(\vec{b}_a - \vec{b}_b)$. Then the evolution equation of the amplitude in rapidity, $y$, may be written in the symbolic form
\be
\frac{\partial F(y,\vec{k}_t,\vec{b})}{\partial y}~=~\int \frac{d^2k'_t}{\pi k^{'2}_t}~  K(\vec{k}_t,\vec{k}'_t)~F(y,\vec{k}'_t,\vec{b})
\label{eq:evol}
\ee
where $\vec{b}=(\vec{b}_a + \vec{b}_b)/2$ and $ K$ is the known BFKL-like kernel. Note that $\partial y=\partial \ln(1/x)$. The iteration of this equation generates the gluon ladder. This is illustrated in Fig.\ref{fig:ladder2}(b) for the example of a Pomeron-exchange interaction of a proton with a quark-antiquark dipole positioned at $\vec{b}_a$ and $\vec{b}_b$ in the impact parameter plane.  

If we were to take 
\be
 K~=~\Delta ~\delta(\vec{k}_t -\vec{k}'_t)\pi k^2_t,
\ee
then the solution is $F(y)\propto{\rm exp}(\Delta y)$, with $\vec{k}_t$ and $\vec{b}$ frozen during the evolution. Thus, this simple approximation gives $F(Y) \sim e^{\Delta Y} \sim s^\Delta$. In this case the evolution equation has the form $\partial F/\partial y=\Delta F$, where $\Delta$ is the probability to emit new intermediate partons per unit of rapidity; it is analogous to the splitting function of DGLAP evolution. 

However, the important new feature, which we explore in this paper, is the diffusion, or random walk, of $\vec{k}_t$ as we evolve along the ladder. This is necessary in order to have a unified description of soft data together with, for example, high $p_t$ jet production. 
To allow for the ln$\k$ diffusion we use a simplified form of the BFKL kernel,
\be
K= N\Delta\sqrt{\frac{k_t^{'2}}{k_t^2}}\exp\left(-\sqrt{d^2+\ln^2
(k_t^{'2}/k^2_t)}/2\right),
\label{eq:K}
\ee
  which takes into account the main qualitative features of the BFKL approach.  It allows for diffusion in ln$k_t$ space with a diffusion coefficient controled by the parameter $d$ which was chosen to reproduce the diffusion caused by 
the BFKL kernel with resummed next-to-leading ln$(1/x)$ (NLL) corrections \cite{kmrsre}. 
Here, the normalization factor, $N$, is chosen so that $1+\Delta$ is the usual bare Pomeron intercept. 
Note that
BKFL evolution is symmetric in both directions of rapidity, or, equivalently, is symmetric in $k_t$ and $k'_t$, when the amplitude is written in terms of the corresponding symmetric function \cite{BFKL}
\be
\varphi=F \sqrt{\frac{k_t^{'2}}{k^2_t}}\propto 1/\sqrt{k^2_t k_t^{'2}}.
\ee
Thus the square root factor in (\ref{eq:K}) is needed,  since we consider evolution for $F(k_t^{'2})\propto 
 1/k^{'2}_t$, which is clearly not symmetric in $k_t$ and $k'_t$.
Recall that the exponential factor in (\ref{eq:K}) is symmetric.  
 Note also that in the collinear limit $k'_t \ll k_t$ (or $k'_t\gg k_t$)
 the kernel (\ref{eq:K}) has the conventional LO logarithmic behaviour 
 $K\propto k_t^{'2}/k^2_t$.

At this point, it is worth emphasizing, that in the model we are proposing, we do not use the precise LO BFKL result (it is known that the NLL corrections are large), but rather 
the simplified 
form, (\ref{eq:K}), which takes into account the main qualitative features of the BFKL approach. It includes diffusion in ln$k_t$ space with a diffusion coefficient $d$ given in an approximate kernel consistent with resummed next-to-leading ln$(1/x)$ (NLL) corrections. Moreover, the intercept $\Delta$ is tuned to describe the available  `soft' data, and has a value close to that given by the NLL resummed BFKL equation, $\Delta\simeq 0.3$, see Table \ref{tab:A1}.

Now we must include absorptive effects. That is, to allow for the contributions from multi-Pomeron diagrams.
Since the intermediate parton may be absorbed by an
interaction
with the particles (partons) from the wave function of both the beam or target
hadron, we now need to solve the two equations\footnote{The data in the triple-Reggeon domain indicate  very small
 $t$-slopes of all the triple-Reggeon
vertices \cite{LKMR,kkpt}. Indeed, the slopes are consistent with zero. Thus, the size of the
multi-Reggeon vertices are negligible in comparison with the size of the
incoming hadron. For this reason the absorptive corrections (that is, the
exponential factors on the right-hand-side  of (\ref{eq:up}) and (\ref{eq:down}) 
such that the opacities $\Omega_i,\, \Omega_k$ are taken at the same
point in the impact parameter plane $\vec{b}$.}
\be
\frac{\partial F_k(y,\k,\b)}{\partial y}~=~\int \frac{d^2\kk}{\pi k^{'2}_t}~{\rm exp}(-\lambda[\Omega_k(y,\kk)+\Omega_i(y',\kk)]/2)~K(\k,\kk)~F_k(y,\kk,\b).
\label{eq:up}
\ee
\be
\frac{\partial F_i(y',\k,\b)}{\partial y'}~=~\int \frac{d^2\kk}{\pi k^{'2}_t}~{\rm exp}(-\lambda[\Omega_i(y',\kk)+\Omega_k(y,\kk)]/2)~K(\k,\kk)~F_i(y',\kk,\b).
\label{eq:down}
\ee
We evolve (\ref{eq:up}) from $y=y_0$ to $y=Y_k-y_0$, and (\ref{eq:down}) from $y'=y_0$ to $y'=Y_k-y_0$; where $y'=Y_k-y$ and $Y_k={\rm ln}(s/k_t^2)$, recall (\ref{eq:Yk}).

\begin{figure}
\begin{center}
\includegraphics[height=3.5cm]{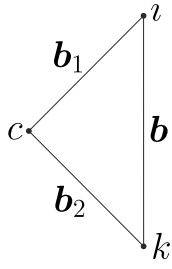}
\caption[*]{\sf The positions of the target $i$, beam $k$ and the intermediate parton $c$ in the impact parameter plane.}
\label{fig:b}
\end{center}
\end{figure}

To calculate $F$ we need to solve these integro-differential evolution equations as a functions of both $y$ and $\vec{k}_t$ for a whole range of fixed values of $\vec{b}$.
 This is done iteratively, giving the solution $F_{ik}$, 
which depends on two indices, that is on the properties of the `beam' and `target' diffractive eigenstates.
Moreover, note that the  $F_{ik}$ now depend on two
 vectors in impact parameter space - the separation ${\vec b}_1$ between
 the position of the intermediate parton $c$ and the beam hadron, and 
the separation ${\vec b}_2$  between $c$ and the target hadron, see Fig. \ref{fig:b}. Recall that the impact parameter of $c$ is that of the `centre', $(\vec{b}_a+\vec{b}_b)/2$, of the Pomeron. The argument $\vec{b}$
in (\ref{eq:up},\ref{eq:down}) now symbolically denotes both
${\vec b}_1$ and ${\vec b}_2$.

We have written (\ref{eq:up}) and (\ref{eq:down}) in terms of the `simpler' form, (\ref{sap-a}), of the absorptive factors, in order to make the evolution equations easier to read. If form (\ref{sap-b}) is used (which is the one adopted in the present model), then the absorptive factor
 $\exp(-\lambda(\Omega_k(y,\vec{k}'_t)+\Omega_i(y',\vec{k}'_t))/2)$ in (\ref{eq:up}) should be replaced by the product
\be
\frac{1-e^{-\lambda\Omega_k(y,\vec{k}'_t)/2}}{\lambda\Omega_k(y,\vec{k}'_t)/2}\cdot 
\frac{1-e^{-\lambda\Omega_i(y',\vec{k}'_t)/2}}{\lambda\Omega_i(y',\vec{k}'_t)/2} ~.
\ee

\section{Expressions for the observables}

Here we summarize how the various amplitudes and cross sections can be calculated in terms of the model proposed above.

\subsection{Elastic amplitude and low-mass diffraction}

\begin{figure}
\begin{center}
\includegraphics[height=4cm]{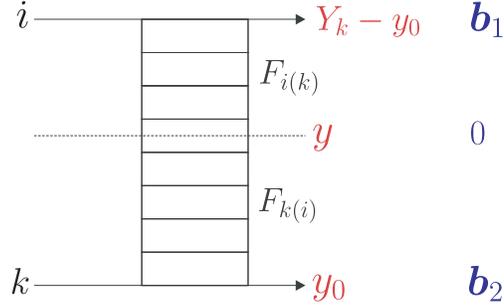}
\caption[*]{\sf The irreducible amplitude $\Omega^{\rm eff}_{ik}({\vec b},Y_k)$ of a high energy interaction.}
\label{fig:Fik}
\end{center}
\end{figure}

To calculate the elastic amplitude we need the $s$-channel two-particle {\it irreducible} amplitudes for the scattering of the various diffractive eigenstates, $i$ and $k$, for given separations  ${\vec b}={\vec b}_1-{\vec b}_2$
 between the incoming protons. These amplitudes, or effective opacities,
 are given by
  \begin{equation}
\Omega^{\rm eff}_{ik}({\vec b},Y_k)=
\int d^2 k_t\frac{\pi^3}2\int F_{ik}({y,\vec k}_t,
{\vec b}_1,{\vec b}_2)~F_{ik}(Y_k-y,{\vec k}_t,{\vec b}_1,{\vec b}_2)~ d^2b_1d^2b_2\delta^{(2)}({\vec b}_1-{\vec b}_2-{\vec b})
 \label{eq:F}
 \end{equation}
 where $Y_k=\ln (s/k^2_t)$, see Fig.~\ref{fig:Fik}.  Note that there is
 no integral\footnote{The integral over $y$ gives the multiplicity.} over
 $y$. The convolution may be calculated at any rapidity $y$, leading to 
the same result. Given this effective `$ik$ eikonal', we can calculate the cross sections (using the analogous relation to (\ref{eq:Tb})). Thus, the total cross section is given by
  \begin{equation}
 \sigma_{\rm tot}=2\sum_{i,k}|a_i|^2|a_k|^2\int\left(1-e^{-\tilde{\Omega}/2}
 \right) d^2b \; ,
 \label{eq:st}
 \end{equation}
where $\tilde{\Omega}$ is shorthand for the effective opacity, (\ref{eq:F}), that is, $\tilde{\Omega} \equiv \Omega^{\rm eff}_{ik}(Y,{\vec b})$. The $a_i$ are the probability amplitudes for the various diffractive eigenstates. That is, for the incoming beam proton we have $|p\rangle=\sum a_i |i\rangle$, and similarly for the incoming target proton. Similarly for the elastic interaction we have
\begin{equation}
 \frac{d\sigma_{\rm el}}{dt}=\frac{1}{4\pi}\left|\int d^2b~ e^{i  
{\vec Q}_t\cdot {\vec b}}\sum_{i,k}|a_i|^2|a_k|^2
 \left(1-e^{-{\tilde\Omega}/2} \right)\right|^2,
 \label{eq:dsel}
 \end{equation}
  \begin{equation}
 \sigma_{\rm el}=\int d^2b\left|\sum_{i,k}|a_i|^2|a_k|^2
 \left(1-e^{-{\tilde \Omega}/2} \right)\right|^2  \; .
 \label{eq:sel}
 \end{equation}
Allowing for the low-mass dissociation of one proton we get
  \begin{equation}
 \sigma_{\rm el+SD}=\int d^2b\sum_i|a_i|^2\left|\sum_{k}|a_k|^2
 \left(1-e^{-{\tilde \Omega}/2} \right)\right|^2  \; .
 \label{eq:selSD}
 \end{equation}

\subsection{High-mass dissociation}
High-mass diffractive dissociation of the beam particle can be written as the elastic scattering of an intermediate parton $c$ due to its absorption on the target, which is described by an amplitude like (\ref{eq:Tb}) but with $\Omega$ replaced by $\lambda\Omega_k(k'_t)$.
Thus, at each impact parameter point ${\vec b}$, and for each value of $k'_t$, the cross section for  single
dissociation is proportional (i) to the elastic $c-k$ cross section
$(1-\exp(-\lambda
\Omega_k(y,k'_t,{\vec b})/2))^2$; (ii) to the probability to find
the intermediate parton $c$ in the interval $dy$, that is
$\Delta\exp(-\lambda
\Omega_i/2-\lambda\Omega_k/2)$; (iii) to the
amplitude $F_i$ of the parton $c$-beam interaction; (iv) to the
gap survival factor $S^2({\vec b})=\exp(-\Omega(Y,{\vec b}))$. The
resulting cross section for high-mass dissociation reads
$$\frac{d\sigma_{\rm SD}}{dy}\,=\, N\int
(1-e^{-\lambda\Omega_k(y,k'_t,{\vec b}_1,{\vec b}_2)/2})\Delta
e^{-\lambda\Omega_i(Y_k-y,k'_t,{\vec b}_1,{\vec b}_2)/2-\lambda\Omega_k(y,k'_t,{\vec b}_1,{\vec b}_2)/2}~~~\times $$
\begin{equation}
\times~~~\frac 12\Omega_k(y,k'_t,{\vec b}_1,{\vec b}_2)F_i(Y_k-y,k'_t,{\vec b}_1,{\vec b}_2)S^2_{ik}(|{\vec b}_1-{\vec b}_2|)dk^{'2}_t
d^2b_1 d^2b_2\; ,
\label{eq:sd}
\end{equation}
where ${\vec b}_1$ (${\vec b}_2$) are the coordinates in
the impact parameter plane with respect to the beam (target) hadron.
The normalisation factor $N$ is specified in (\ref{eq:sdf}).
The eikonal gap survival probability\footnote{Strictly speaking, when calculating the
gap survival probability in each particular case, we only have to account for
the possibility of rescattering which
produces secondaries within the gap interval. That is, in
(\ref{eq:s}) we should not put the whole irreducible amplitude
$\tilde{\Omega}_{ik}({\vec b})$, but, instead, part of it; since the contribution from the processes
with a gap in the same (or a larger) rapidity interval does not change
 qualitatively the structure of the diffractive dissociation
event. In the present computations we neglect this
effect. This means that actually the gap survival probabilities, and
the true cross sections of diffractive dissociation, should be a bit
larger.}
\begin{equation}
S^2_{ik}({\vec b})=\exp(-\tilde{\Omega}_{ik}^{\rm eff}({\vec b}))\; .
\label{eq:s}
\end{equation}

First, we consider hypothesis (\ref{eq:(a)}) for the multi-Pomeron vertices $g^n_m$. Accounting for the different Good-Walker eigenstates, and integrating over the transverse momentum $k'_t$ of parton $c$, we obtain
$$\frac{M^2d\sigma_{\rm SD}}{dM^2}\,=\sum_i|a_i|^2\frac{\pi^3}2\int d k_t^{'2}\int
\left|\sum_k|a_k|^2\sqrt{T_{ik}(y,k'_t,{\vec b}_1,{\vec b}_2)
\rho_{ik}(y,k'_t,{\vec b}_1,{\vec b}_2))}S_{ik}(|{\vec b}_1-{\vec b}_2|)\right|^2~~\times
$$
\begin{equation}
\times~~\frac 12\Omega_k(y,k'_t,{\vec b}_1,{\vec b}_2)F_i(Y_k-y,k'_t,{\vec b}_1,{\vec b}_2)d^2b_1d^2b_2\, ,
\label{eq:sdf}
\end{equation}
where the parton density
\begin{equation}
\rho_{ik}=\Delta
e^{-\lambda(\Omega_k(y,k'_t,b)+
\Omega_i(Y_k-y,k'_t,b))/2},
\label{eq:rho}
\end{equation}
and the elastic
 $c-k$ amplitude
\begin{equation}
T_{ik}(y,k'_t,{\vec b}_1,{\vec b}_2)=
\left(1-e^{-\lambda\Omega_k(y,k'_t,{\vec b}_1,{\vec b}_2)/2}\right)\; .
\label{eq:T}
\end{equation}

In an analogous way,  the expressions for the $t$-slope of high-mass diffractive 
dissociation, for the cross section of Central Exclusive Production etc., can be obtained from corresponding formulae given in \cite{KMRnns1}.

For hypothesis (\ref{eq:(b)}) for the multi-Pomeron vertices $g^n_m$, we find
 use of the AGK rules \cite{AGK} shows 
that expression (\ref{eq:sdf}) takes the form
$$\frac{M^2d\sigma_{\rm SD}}{dM^2}\,=\sum_i|a_i|^2\frac{\pi^3}2\int d k_t^{'2}\int
\left|\sum_k|a_k|^2T_{ik}(y,k'_t,{\vec b}_1,{\vec b}_2)
\sqrt{\overline{\rho}_{ik}(y,k'_t,{\vec b}_1,{\vec b}_2))}S_{ik}(|{\vec b}_1-{\vec b}_2|)\right|^2~~\times
$$
\begin{equation}
\times~~F_i(Y_k-y,k'_t,{\vec b}_1,{\vec b}_2,Y_k-y)d^2b_1d^2b_2\, ,
\label{eq:sdf1}
\end{equation}
with 
\begin{equation}
\overline{\rho}_{ik}=\frac{\Delta}{\lambda}
(1-e^{-\lambda\Omega_i(Y_k-y,k'_t,b))/2})/
(\lambda\Omega_i(Y_k-y,k'_t,b))/2).
\label{eq:rho1}
\end{equation}

\subsection{Hadron $p_t$ spectra \label{sec:B3}}
From the AGK cutting rules \cite{AGK}, it follows that the inclusive (gluon) jet distribution is given just by the irreducible amplitude
\begin{equation}
\frac{d\sigma}{dydq^2_t}=
\sum_{i,k}|a_i|^2|a_k|^2\int_{q_t}\frac{\pi^3 dk^{'2}_t}{2q^2_t}
\rho_{ik}(y,k_t,b)\underline{F}_{ik}(y,{\vec k}_t,
{\vec b}_1,{\vec b}_2)~\underline{F}_{ik}(Y_k-y,{\vec k}_t,{\vec b}_1,{\vec b}_2)~ d^2b_1d^2b_2.
 \label{eq:pt}
\end{equation}
Here, following the AGK rules \cite{AGK}, we use the gluon densities, $\underline{F}$ of (\ref{eq:glutt}), calculated without the absorption caused by the diagrams which cross the rapidity of the emitted (inclusive) gluon\footnote{That is, without the absorptive effects in target opacity which arise from the beam side.}; and
 account for the fact that any BFKL ladder, as in Fig.~\ref{fig:Pladder}, with a transverse
momentum of the $t$-channel gluon $k'_t>q_t$ may emit a `soft' gluon $q_t$ with the same vertex squared factor $|V|^2\propto 1/q^2_t$. Therefore 
(\ref{eq:pt}) contains an integral over $k'_t$.

To obtain the hadron $p_t$ spectra this distribution should be convoluted with the gluon fragmentation function $D_g(z,q^2_t)$, where $z=p_t/q_t$.
We assume a simplified gluon fragmentation function 
\begin{equation}
D_g(z,q_t)=\frac Nz\cdot (1-z)^{5+\delta}~~~~~{\rm with}~~~\delta=\int_{q_0}^{q_t}N_c\alpha_s(k')d\ln k^{'2}/\pi
\label{eq:D} 
\end{equation}
where the power $\delta$ accounts for the main  (double-logarithmic) effect of the scaling violation in the large $z$ region. The normalization factor $N$ is fixed by energy-momentum conservation.
In the relevant region of $z$ this function is in satisfactory agreement
 with the available data on gluon jet fragmentation \cite{gjf}.
 
 The results shown in Fig.~\ref{fig:pT} do not account for the secondaries  coming from the
hadronization of (the LUND or other hadronization) colour strings with small $q_t$. Thus the deficiency of the prediction at low $p_t$, seen in  Fig.~\ref{fig:pT}, may be removed by the particles produced via hadronization. To reproduce this contribution we need an explicit Monte Carlo realization of the model.

\section*{Acknowledgements}
MGR would like to thank the IPPP at the University of Durham for hospitality. This work was supported by the grant RFBR 11-02-00120-a
and by the Federal Program of the Russian State RSGSS-65751.2010.2.


\thebibliography{}

\bibitem{gmc}  A.~Buckley, J.~Butterworth, S.~Gieseke {\it et al.},
 [arXiv:1101.2599 [hep-ph]].




\bibitem{P81} T.~Sjostrand, S.~Mrenna and P.Z.~Skands,
  Comput.\ Phys.\ Commun.\  {\bf 178}, 852 (2008)
  [arXiv:0710.3820 [hep-ph]].


\bibitem{book} for a recent detailed review see
V.S.~Fadin, B.L.~Ioffe and L.N.~Lipatov,
Quantum Chromodynamics
(Camb. Univ. Press, 2010).




\bibitem{RFT}V.N.~Gribov, Sov. Phys. JETP {\bf 26}, 414 (1968).

\bibitem{bkk} for a recent review and references see
 K.G.~Boreskov, A.B.~Kaidalov and O.V.~Kancheli,
  Phys.\ Atom.\ Nucl.\  {\bf 69} (2006) 1765
  [Yad.\ Fiz.\  {\bf 69} (2006) 1802].


\bibitem{BK}
  I.~Balitsky,
  Nucl.\ Phys.\  {\bf B463}, 99 (1996)
  [arXiv:hep-ph/9509348];\\ 
Y.V.~Kovchegov,
  Phys.\ Rev.\  {\bf D60}, 034008 (1999)
  [arXiv:hep-ph/9901281].



\bibitem{atlas}  G.~Aad {\it et al.}  [ATLAS Collaboration],
  arXiv:1012.0791 [hep-ex];   arXiv:1012.5104 [hep-ex]. 


\bibitem{cms} V.~Khachatryan {\it et al.}  [CMS Collaboration],
  Phys.\ Rev.\ Lett.\  {\bf 105}, 022002 (2010)
  [arXiv:1005.3299 [hep-ex]].


\bibitem{regge} P.D.B. Collins, {\it Regge theory and high energy physics}, (Cambridge Univ. Press, 1977); \\
A.C. Irving and R.P. Worden, Phys. Rept. {\bf 34}, 117 (1977).

\bibitem{DL}
 A.~Donnachie and P.V.~Landshoff,
  Phys.\ Lett.\  B {\bf 296}, 227 (1992)
[arXiv:hep-ph/9209205].


\bibitem{M-K} A.H. Mueller, Phys. Rev. {\bf D2}, 2963 (1970);\\
 O. Kancheli, JETP Lett. {\bf 11}, 267 (1970).

\bibitem{AGK}V.A.~Abramovsky, V.N.~Gribov and O.V.~Kancheli, Sov. J.
Nucl. Phys. {\bf 18}, 308 (1973).


\bibitem{cer} V.A.~Khoze, A.D.~Martin and M.G.~Ryskin,
  Nucl.\ Phys.\ Proc.\ Suppl.\  {\bf 99B}, 213 (2001)
  [arXiv:hep-ph/0011319].


\bibitem{BL} I.I.~Balitsky and L.N.~Lipatov,
  JETP Lett.\  {\bf 30} (1979) 355
  [Pisma Zh.\ Eksp.\ Teor.\ Fiz.\  {\bf 30} (1979) 383];\\
``Regge Processes In Nonabelian Gauge Theories. (In Russian),''
{\it  in Proceedings, Physics Of Elementary Particles, Leningrad 1979, p109-149}.

\bibitem{BFKL86} L.N. Lipatov, Sov.\ Phys.\ JETP {\bf 63}, 904 (1986)
 [Zh.\ Eksp.\ Teor.\ Fiz.\  {\bf 90}, 1536 (1986)].

\bibitem{LKMR} E.G.S. Luna, V.A. Khoze, A.D. Martin and M.G. Ryskin, Eur. Phys. J. {\bf C59}, 1 (2009).

\bibitem{deOliv}
  E.G.~de Oliveira, A.D.~Martin and M.G.~Ryskin,
  Phys.\ Lett.\  {\bf B695}, 162 (2011)
  [arXiv:1010.1366 [hep-ph]].

\bibitem{bor} P. Grassberger, K. Sundermeyer, Phys. Lett. {\bf B77}, 220 (1978); \\
K. Boreskov, hep-ph/0112325.


\bibitem{KMRns}  M.G.~Ryskin, A.D.~Martin and V.A.~Khoze,
  Eur.\ Phys.\ J.\  {\bf C54}, 199 (2008)
[arXiv:0710.2494 [hep-ph]].

  


\bibitem{shuv} M.G.~Ryskin, A.D.~Martin, V.A.~Khoze and A.G.~Shuvaev,
  J.\ Phys.\  {\bf G36}, 093001 (2009)
  [arXiv:0907.1374 [hep-ph]].


\bibitem{KMRnns1}  M.G.~Ryskin, A.D.~Martin and V.A.~Khoze,
  Eur.\ Phys.\ J.\  {\bf C60}, 249 (2009)
  [arXiv:0812.2407 [hep-ph]].


\bibitem{GLMM} E. Gotsman, E. Levin, U. Maor and J.S. Miller, Eur. Phys. J. {\bf C57}, 689 (2008);\\
 E.~Gotsman, E.~Levin and U.~Maor,
  arXiv:1010.5323 [hep-ph].



\bibitem{bfklresum} M. Ciafaloni, D. Colferai and G. Salam, Phys. Rev. {\bf D60}, 114036 (1999); \\
G. Salam, JHEP {\bf 9807}, 019 (1998); Act. Phys. Pol. {\bf B30}, 3679 (1999).

\bibitem{kmrsre}V.A.~Khoze, A.D.~Martin, M.G.~Ryskin and W.J. Stirling, Phys. Rev. {\bf D70}, 074013 (2004).


\bibitem{theta} L.V. Gribov, E. Levin and M.G. Ryskin, Phys. Rept. {\bf 100}, 1 (1983) [Yad. Fiz. {\bf 35}, 1278 (1982)];\\
J. Bartels and M. W\"{u}sthoff, Z. Phys. {\bf C66}, 157 (1995).


\bibitem{GW} M.L. Good and W.D. Walker, Phys. Rev. {\bf 120}, 1857 (1960).


\bibitem{CDFhm} F. Abe {\it et al.}, [CDF collaboration]  Phys. Rev. {\bf D50}, 5535 (1994).

\bibitem{dino}
 K. Goulianos and J. Montanha, Phys. Rev. {\bf D59} 114017 (1999).


\bibitem{PhysRep} K.~Goulianos,
  Phys.\ Rept.\  {\bf 101} (1983) 169.

\bibitem{data} F.Abe et al. [CDF Collaboration], Phys. Rev. Lett. {\bf 61}, 1819 (1988).

\bibitem{KMR} V.A.~Khoze, A.D.~Martin and M.G.~Ryskin,
  Eur.\ Phys.\ J.\   {\bf C23}, 311 (2002)
  [arXiv:hep-ph/0111078];\\
  V.A.~Khoze, A.D.~Martin, M.G.~Ryskin and W.J.~Stirling,
  Eur.\ Phys.\ J.\  {\bf C35}, 211 (2004)
  [arXiv:hep-ph/0403218];\\
V.A.~Khoze, A.D.~Martin, M.G.~Ryskin and W.J.~Stirling,
  Eur.\ Phys.\ J.\  {\bf C38}, 475 (2005)
  [arXiv:hep-ph/0409037];\\
L.A.~Harland-Lang, V.A.~Khoze, M.G.~Ryskin and W.J.~Stirling,
  Eur.\ Phys.\ J.\   {\bf C69} (2010) 179
  [arXiv:1005.0695 [hep-ph]];\\
L.A.~Harland-Lang, V.A.~Khoze, M.G.~Ryskin and W.J.~Stirling,
  arXiv:1011.0680 [hep-ph].

\bibitem{CERN-ISR} L.~Baksay {\it et al.}, Phys.\ Lett.\ {\bf B53}, 484 (1975); \\
R.~Webb {\it et al.}, Phys.\ Lett.\ {\bf B55}, 331 (1975); \\
L.~Baksay {\it et al.}, Phys.\ Lett.\ {\bf B61}, 405 (1976); \\
H.~de Kerret {\it et al.}, Phys.\ Lett.\ {\bf B63}, 477 (1976); \\
G.C.~Mantovani {\it et al.}, Phys.\ Lett.\ {\bf B64}, 471 (1976).
\bibitem{angr}  A.A. Anselm and V.N. Gribov, Phys.\ Lett.\ {\bf B40}, 487 (1972).
\bibitem{KMRsoft}
V.A.~Khoze, A.D.~Martin and M.G.~Ryskin,
  Eur.\ Phys.\ J.\   {\bf C18}, 167 (2000)
  [arXiv:hep-ph/0007359].
\bibitem{ost} 
  S.~Ostapchenko,
  Phys.\ Rev.\   {\bf D81}, 114028 (2010)
  [arXiv:1003.0196 [hep-ph]];\\
arXiv:1010.1869 [hep-ph].
\bibitem{fsc} M.~Albrow {\it et al.}  [USCMS Collaboration],
  JINST {\bf 4}, P10001 (2009)
  [arXiv:0811.0120 [hep-ex].


\bibitem{MSTW} A.D. Martin, W.J. Stirling, R.S. Thorne and G. Watt, Eur. Phys. J. {\bf C63}, 189 (2009).


\bibitem{MRW} A.D.~Martin, M.G.~Ryskin and G.~Watt,
  Phys.\ Lett.\   {\bf B644}, 131 (2007)
  [arXiv:hep-ph/0609273].

\bibitem{KMRnns2} M.G. Ryskin, A.D. Martin and V.A. Khoze, Eur. Phys. J. {\bf C60}, 265 (2009).
\bibitem{GLM11} E. Gotsman, E. Levin and U. Maor, arXiv:1101.5816 [hep-ph].



\bibitem{KPT}A.B. Kaidalov, L.A. Ponomarev and K.A. Ter-Martirosyan, Sov. J. Nucl.     
Phys. {\bf 44} (1986) 468.

\bibitem{UHECR} D. d'Enterria, R. Engel, T. Pierog, S. Ostapchenko and K. Werner, arXiv:1101.5596 [astro-ph].

\bibitem{Krauss} F. Krauss et al., in preparation.

\bibitem{BFKL}   V.S.~Fadin, E.A.~Kuraev, and L.N.~Lipatov,
Phys. Lett.  {\bf B60}, 50  (1975); \\
E.A.~Kuraev, L.N.~Lipatov, and V.S.~Fadin, Zh. Eksp. Teor. Fiz.
{\bf 71}, 840 (1976) [Sov. Phys. JETP {\bf 44}, 443 (1976)]; {\it
ibid.} {\bf 72}, 377 (1977) [{\bf 45}, 199 (1977)];\\
I.I.~Balitsky and L.N.~Lipatov, Yad. Fiz. {\bf28}, 1597 (1978)
[Sov. J. Nucl. Phys. {\bf28}, 822 (1978)].
%

\bibitem{kkpt} A.B.~Kaidalov, V.A.~Khoze, Yu.F.~Pirogov and N.L.~Ter-Isaakyan,
  Phys.\ Lett.\   {\bf B45} (1973) 493.

\bibitem{gjf}
  G.~Abbiendi {\it et al.}  [OPAL Collaboration],
  Eur.\ Phys.\ J.\  {\bf C11}, 217 (1999)
  [arXiv:hep-ex/9903027];\\
  S.~Albino, B.A.~Kniehl, G.~Kramer and W.~Ochs,
  Phys.\ Rev.\   {\bf D73}, 054020 (2006)
  [arXiv:hep-ph/0510319].
\end{document}